\documentclass[useAMS,usenatbib]{mn2e}
\usepackage{amsmath}
\usepackage{graphicx}
\usepackage{epstopdf}
\usepackage{amssymb}
\usepackage{extarrows}
\usepackage{color}
\newcommand{\jvec}{\mathbf{j}}
\newcommand{\evec}{\mathbf{e}}
\newcommand{\rvec}{\mathbf{r}}
\newcommand{\Ham}{\mathcal{H}}
\newcommand{\vvec}{\hat{\mathbf{v}}}
\newcommand{\uvec}{\hat{\mathbf{u}}}
\newcommand{\nvec}{\hat{\mathbf{n}}}
\newcommand{\oct}{\mathrm{Oct}}
\newcommand{\K}{\mathrm{Quad}}
\newcommand{\gr}{\mathrm{GR}}
\newcommand{\tide}{\mathrm{Tide}}
\newcommand{\rot}{\mathrm{Rot}}
\newcommand{\m}{\mathrm{max}}
\newcommand{\mi}{\mathrm{min}}

\newcommand{\au}{\mathrm{AU}}
\newcommand{\li}{\mathrm{lim}}

\newcommand{\extra}{\mathrm{extra}}
\def\e1{e_1^2}
\def\I{i_{\mathrm{tot}}}
\def\G1{L_1\sqrt{1-e_1^2}}

\title[Triple systems with short range forces]{Suppression of extreme orbital evolution in triple systems with short range forces}

\author[Liu, Mu\~noz \& Lai]{Bin Liu$^{1,2}$\thanks{E-mail: bl559@cornell.edu},
Diego J. Mu\~noz$^{2}$
and Dong Lai$^{2}$\\
$^{1}$ Center for Astrophysics, University of Science and Technology of China, Hefei, Anhui 230026, People's Republic of China\\
$^{2}$ Center for Space Research, Department of Astronomy, Cornell University, Ithaca, NY 14853, USA
}
\begin{document}

%\date{Accepted . Received ; in original form }

\pagerange{\pageref{firstpage}--\pageref{lastpage}} \pubyear{2014}

\maketitle

\label{firstpage}

\begin{abstract}
The Lidov-Kozai (LK) mechanism plays an important role in the secular evolution of 
many hierarchical triple systems. The standard LK mechanism consists of large-amplitude oscillations in eccentricity and inclination of a binary subject to the quadrupole potential from an outer perturber. Recent work has shown that when the octupole terms are included in the potential, the inner binary can reach more extreme eccentricities as well as undergo orientation flips. It is known that pericenter precessions due to short-range effects, such as General Relativity and tidal and rotational distortions, can limit the growth of eccentricity and even suppress standard (quadrupolar) LK oscillations, but their effect on the octupole-level LK mechanism has not been fully explored. In this paper, we systematically study how these short-range forces affect the extreme orbital behaviour found in octupole LK cycles. In general, the influence of the octupole potential is confined to a range of initial mutual inclinations $i_\mathrm{tot}$ centered around 90$^\circ$ (when the inner binary mass ratio is $\ll1$), with this range expanding with increasing octupole strength. We find that, while the short-range forces do not change the width and location of this "window of influence", they impose a strict upper limit on the maximum achievable eccentricity. This limiting eccentricity can be calculated analytically, and its value holds even for strong octupole potential and for the general case of three comparable masses. Short-range forces also affect orbital flips, progressively reducing the range of $i_\mathrm{tot}$ within which flips are possible as the intensity of these forces increases.
\end{abstract}


\section{Introduction}

Three-body systems are ubiquitous in astrophysics, appearing in a wide
range of configurations and scales, from planet-satellite systems to
black holes in dense stellar clusters.  Although the gravitational
three-body problem is in general non-integrable, a hierarchical system
(i.e., triple configuration consisting of an inner binary orbited by a
distant companion) can be simplified by retaining the lowest orders in
the multipole expansion of the interaction potentials.  In this case,
the triple system is represented by two nested binary systems (an
``inner binary" and an ``outer binary"), with the corresponding
orbital elements evolving on secular timescales due to mutual
interactions.

\citet{Lidov} and \citet{Kozai} discovered that when the mutual
inclination angle between the inner and outer binaries is sufficiently
high, the time-averaged tidal gravitational force from the outer
companion can induce large-amplitude oscillations in the eccentricity
and inclination of the inner binary%%%%%%%%%%%%%%%
\footnote{Lidov considered the long-term
evolution of satellite orbits under the perturbation of the Moon,
while Kozai studied the evolution of asteroid orbits under the
perturbation of Jupiter.%%%%%%%%%%%%%%%%%. 
}. In recent years, numerous works have shown
that Lidov\---Kozai oscillations could play an important role in the
formation and evolution of various astrophysical systems. Examples
include:
(i) The formation of close stellar binaries, including those
containing compact objects
\citep[e.g.,][]{MS,KEM,Eggleton,FT,PF,ST,Smadar 2014};
(ii) The excitation of eccentricities of exoplanet
systems
\citep[e.g.,][]{Holman,Innanen,Mazeh}
and the formation of hot Jupiters through high-eccentricity
migration 
{\citep[e.g.,][]{WM,FT,Correia,Smadar 2012,Subo Dong,sto14,Petrovich}};
(iii) The production of
Type Ia supernovae from white-dwarf binary mergers
\citep[e.g.,][]{Thompson,PMT}
or
direct collisions \citep[e.g.,][]{Katz,Kushnir};
(iv) The properties of irregular satellites (particularly their
inclination distribution relative to the ecliptic)
of giant planets in the solar system
\citep[e.g.,][]{Carruba,Nesvorny};
(v) The formation and merger of (stellar and supermassive)
black hole binaries at the centers globular clusters or
galaxies
\citep[e.g.,][]{Blaes,MH,Wen,AMM}.

The simplest Lidov\---Kozai mechanism involves a test mass (``planet'')
orbiting a primary body (``star'') perturbed by an external companion,
with the interaction potential truncated to the quadrupole order. In
this test-mass, quadrupole approximation, the projected angular
momentum (along the external binary axis) of the planet is conserved.
If the influence of other short-range forces (SRFs) is negligible,
the maximum eccentricity achieved by the inner binary (for an
initially very small eccentricity) during the Lidov\---Kozai oscillation
is given by 
\begin{equation}\label{eq:EMAX}
e_\m=\Big(1-\frac{5}{3}\cos^2i_0\Big)^{1/2}
\end{equation}
where $i_0$ is the initial inclination angle of the two orbits.
Thus Lidov\---Kozai oscillation requires $i_0$ to lie between
$\cos^{-1}\sqrt{3/5}\simeq 39^\circ$ and $141^\circ$.
It has been recognized that the Lidov\---Kozai cycles can be
suppressed by other short-range effects that induce periapse
precession of the inner binary, including the precessions due to
General Relativity (GR), rotational bulge and tidal distortion
{\citep[e.g.,][]{Holman, Subo Dong}}.
The suppression arises because these additional precessions
tend to destroy the near $1:1$ resonance between the longitude of
the periapse $\omega$ and the longitude of the ascending node
$\Omega$ required for eccentricity excitation. Thus, the maximum eccentricity
can be reduced from the ``pure'' Lidov\---Kozai value (Equation \ref{eq:EMAX}).

It has also been recognized that high-order expansion of the
interaction potential can lead to a much richer dynamical behaviour of
hierarchical triples than the simplest Lidov\---Kozai oscillation based
on the test-mass, quadrupole approximation.
\citet{Harrington,Marchal,Krymolowski,Ford,Blaes}
have derived the orbit-averaged Hamiltonian to
octupole order and used the resulting equation of motion to explore
some aspects of the evolution of triples. Unlike the pure quadrupole
case, the projected angular momentum of the inner binary (even in the
test-mass limit) is no longer constant when the octupole potential is
included (the octupole potential is nonzero when the outer binary is
eccentric and the components of the inner binary have different
masses). Therefore, the secular dynamics of triples is generally not
integrable in the octupole order and may lead to chaos
\citep[e.g.,][]{Li 2014}.
Recent works have examined the rich dynamical behaviour of
such ``eccentric'' Lidov\---Kozai mechanism, either numerically
\citep[e.g.,][]{Smadar 2011,Smadar 2013b,Teyssandier, Li 2013}
or semi-analytically
\citep[e.g.,][]{Katz PRL,LS}, and explored their implications for the
formation of hot Jupiters and the resulting spin-orbit misalignments
\citep[e.g.,][]{Smadar 2011,Smadar 2012,Petrovich}.

The works cited above have revealed two important consequences of the
``eccentric'' Lidov\---Kozai mechanism: (i) The eccentricity of the inner
binary can be driven to extreme value ($1-e\sim 10^{-6}$) even for
``modest'' initial orbital inclinations; (ii) The inner orbit can flip
and come retrograde relative to the outer orbit.  These two effects
are related, as orbital flip is often associated with extreme
eccentricity. Since the precession of periapse due to short-range
forces is strongly dependent on eccentricity, it is not clear
to what extent the extreme eccentricity can be realized in realistic
situations. While short-range effects were included in some population
synthesis calculations for the formation of hot Jupiters
\citep[e.g.,][]{Smadar 2012,Petrovich}, a systematic study of the short-range force effects
on eccentric Lidov\---Kozai mechanism is currently lacking.

In this paper, by running a sequence of numerical integrations, we
study how SRFs affect the evolution of the inner binary
(with and without the test-mass approximation), including the
interaction potential up to the octupole order.
Combining with various analytical considerations,
we characterize the parameters space systematically to understand
how the maximum eccentricity is modified by the SRFs.

Our paper is organized as follows. In Section~2, we derive the secular
equations of motion up to the octupole order using a vectorial
formalism.  In Section~3, we provide a brief overview of short-range
effects, estimating the maximum eccentricity allowed by the presence
of various SRFs. In Section~4, we describe our numerical
integrations, carried out over a range of parameters for triple
systems consisting of a star-planet binary and an outer stellar
companion. In Section~5, we extend our analysis to triple systems in
which all components have comparable masses.  We summarize our main
results in Section~6.

%%%%%%%%%%%%%%%%%%%%%%%%%%%%%%%%%%%%%%%%%%%%%%%%%%%%%%%%%%
\section[]{Evolution of triple systems in the secular approximation}

In a hierarchical triple system, two bodies of masses $m_0$ and $m_1$
orbit each other (with semimajor axis $a_1$) while a third body of mass $m_2$
orbits the center mass of the inner bodies ($m_0$ and $m_1$) on a wider orbit (with semimajor axis $a_2$).
The complete Hamiltonian of the system
can then be written as the sum of the individual Hamiltonians of the inner and
outer orbits plus an interaction potential $\Phi$ \citep[e.g.,][]{Harrington}:
\begin{equation}\label{eq:Ham}
\begin{split}
\Ham=&~~\Ham_1+\Ham_2 + \Phi\\
=&-\frac{G m_0m_1}{2a_1}-\frac{G m_2(m_0+m_1)}{2a_2} \\
& - \frac{G}{a_2} \sum_{l=2}^\infty\left(\frac{a_1}{a_2}\right)^l M_l \left(\frac{|\mathbf{r}_1|}{a_1}\right)^l\left(\frac{a_2}{|\mathbf{r}_2|}\right)^{l+1}P_l(\cos{\theta})~.
\end{split}
\end{equation}
where $\mathbf{r_1}$ is the instantaneous separation vector
 between the inner masses $m_0$ and $m_1$,
$\mathbf{r_2}$ is the instantaneous separation vector between $m_2$ and
center of mass of $m_0$ and $m_1$, and $\theta$  is the angle
between $\rvec_1$  and $\rvec_2$. In Equation~(\ref{eq:Ham}),
$P_l(x)$ is the Legendre polynomial of degree $l$ and $M_l$ is a coefficient  that
depends on $m_0$, $m_1$, $m_2$ and $l$.

If $m_2$ is sufficiently distant (i.e., $a_2\gg a_1$),
only the smallest values of $l$ contribute significantly to $\Phi$, and
the coupling term is weak such that the inner and outer
Keplerian orbits change very slowly (on timescales much longer
than their orbital periods). In this regime, the secular approximation
is valid, meaning that, the system can be adequately described by two
slowly evolving Keplerian orbits,
while the short timescale behaviour of the
three individual trajectories is irrelevant \citep[e.g.,][]{Marchal}.

This perturbative method has been used extensively to study three-body systems
up to quadrupole ($l=2$)\citep[e.g.,][]{Kozai,Lidov} and octupole ($l=3$) orders
\citep[e.g.,][]{Ford, LS, Smadar 2013b}.  Most of these studies have used
the classical perturbation methods of celestial mechanics, based on an orbital-element
formulation of the Hamiltonian system. In the following, we present the secular evolution equations
to the octupole order using a geometric (vectorial) formalism \citep[e.g.][]{Tremaine,Correia,TY},
and confirm via angular projections that they are equivalent to Hamilton's equations for the orbital elements.

%%%%%%%%%%%%%%%%%%%%%%%%%%%%%%%
\subsection{Equations of motion in vector form}
In vector form, the instantaneous position $\rvec$ of a body in Keplerian motion can
be written  as
\begin{equation}\label{eq:keplerian}
\rvec=r(\cos\!{f}\uvec+\sin\!{f}\vvec)
\end{equation}
 with $r=a(1-e^2)/(1+e\cos\!f)$, where
$a$, $e$ and $f$ are the semimajor axis, eccentricity and true anomaly,
respectively. The orthogonal unit vectors $\uvec$ and $\vvec$ define the orbital plane,
where $\uvec$ points in the direction of pericenter (i.e., at $f=0$).
A third unit vector $\nvec$, pointed in the direction of the orbital angular momentum,
completes an orthonormal triad, $\uvec\times\vvec=\nvec$. Alternatively, it is often useful to work in
terms of the dimensionless angular momentum vector $\jvec$ and the eccentricity
vector $\evec$:
\begin{equation}\label{eq:defineje}
\jvec=\sqrt{1-e^2}\nvec~,~~~~~~~~~~~~~~\\
\evec=e\uvec~.
\end{equation}
where $\jvec$ and $\evec$ satisfy $\jvec\cdot\evec=0$ and
$\jvec^2+\evec^2=1$.

Truncating the interaction potential (Equation \ref{eq:Ham}) at the $l=3$ order,
we write $\Phi=\Phi_{\K}+\Phi_{\oct}$, where the quadrupole term is
\begin{equation}\label{eq:PK}
\Phi_{\K}
=-\frac{Gm_0m_1m_2}{(m_0+m_1)r_2}\left[\frac{3}{2}\frac{(\rvec_1\cdot\rvec_2)^2}{r_2^4}-\frac{r_1^2}{2r_2^2}\right],
\end{equation}
and the octupole term is
\begin{equation}\label{eq:PO}
\Phi_{\oct}
=-\frac{Gm_0m_1m_2(m_0-m_1)}{(m_0+m_1)^2r_2}\left[\frac{5}{2}\frac{(\rvec_1\cdot\rvec_2)^3}{r_2^6}-
\frac{3}{2}\frac{r_1^2(\rvec_1\cdot\rvec_2)}{r_2^4}\right]~~,
\end{equation}
and where the position vectors $\rvec_1$ and $\rvec_2$ track
two different Keplerian orbits (Equation \ref{eq:keplerian}) of
orbital elements $a_1$ and $e_1$ (inner) and $a_2$ and $e_2$ (outer),
which are oriented in space by the triads $(\uvec_1,\vvec_1,\nvec_1)$
and $(\uvec_2,\vvec_2,\nvec_2)$, respectively.

The next step is to filter out the high-frequency behaviour by time-averaging
 the quadrupole and octupole potentials twice: over the inner orbital period
 and the outer orbital period. Using a standard
 averaging procedure \citep[e.g.][]{TY}, we find that
the double averaged quadrupole potential is given by
\begin{equation}\label{eq:APK}
\langle\Phi_{\K}\rangle=\frac{\mu_1\Phi_0}{8}\Big[
1-6e_1^2-3(1-e_1^2)(\nvec_1\cdot\nvec_2)^2+15e_1^2(\uvec_1\cdot\nvec_2)^2\Big]~~,
\end{equation}
where $\mu_1=m_0m_1/(m_0+m_1)$ is the reduced mass of the inner orbit.
The double-averaged octupole potential is
\begin{equation}\label{eq:APO}
\begin{split}
\langle\Phi_{\oct}\rangle=&\frac{15\mu_1\Phi_0\varepsilon_{\oct}}{64}
\Bigg\{e_1(\uvec_1\cdot\uvec_2)\Big[8e_1^2-1-35e_1^2(\uvec_1\cdot\nvec_2)^2\\
&~~~~~~~~~~~~~~~~~~~~~~~~~~~~~~+5(1-e_1^2)(\nvec_1\cdot\nvec_2)^2\Big]\\
&~~~~~~~~~~+10e_1(1-e_1^2)(\uvec_1\cdot\nvec_2)
(\nvec_1\cdot\uvec_2)(\nvec_1\cdot\nvec_2)\Bigg\}~~.
\end{split}
\end{equation}

In Equations~(\ref{eq:APK}) and (\ref{eq:APO}) we have defined the coefficients
\begin{equation}
\Phi_0\equiv\frac{Gm_2a_1^2}{a_2^3(1-e_2^2)^{3/2}}~
\end{equation}
and
\begin{equation}\label{eq:C}
\varepsilon_{\oct}\equiv\frac{m_0-m_1}{m_0+m_1}\frac{a_1}{a_2}\frac{e_2}{1-e_2^2}~,
\end{equation}
where the magnitude of $\varepsilon_{\oct}$ quantifies the  importance of the octupole term
relative to the quadrupole term.

In terms of the averaged potentials, the equations of motion for the
orbital vectors $\jvec_1$, $\evec_1$, $\jvec_2$ and $\evec_2$ (defined
as in Equation~\ref{eq:defineje} for the inner and outer orbits) are
\begin{eqnarray}
~~~~~&&\frac{d{\jvec_1}}{dt}=-\frac{1}{L_1}\left(\frac{}{}\jvec_1\times\nabla_{\jvec_1}\langle\Phi\rangle
+\evec_1\times\nabla_{\evec_1}\langle\Phi\rangle \right),\label{eq:Gj1}\\
~~~~~&&\frac{d{\evec_1}}{dt}=-\frac{1}{L_1}\left(\frac{}{}\jvec_1\times\nabla_{\evec_1}\langle\Phi\rangle
+\evec_1\times\nabla_{\jvec_1}\langle\Phi\rangle \right),\label{eq:Ge1}\\
~~~~~&&\frac{d{\jvec_2}}{dt}=-\frac{1}{L_2}\left(\frac{}{}\jvec_2\times\nabla_{\jvec_2}\langle\Phi\rangle
+\evec_2\times\nabla_{\evec_2}\langle\Phi\rangle \right),\label{eq:Gj2}\\
~~~~~&&\frac{d{\evec_2}}{dt}=-\frac{1}{L_2}\left(\frac{}{}\jvec_2\times\nabla_{\evec_2}\langle\Phi\rangle
+\evec_2\times\nabla_{\jvec_2}\langle\Phi\rangle \right).\label{eq:Ge2}
\end{eqnarray}
Here, $L_1$  and $L_2$  are
\begin{eqnarray}
~~~~~&&L_1=\mu_1\sqrt{G(m_0+m_1)a_1},\label{eq:L1}\\
~~~~~&&L_2=\mu_2\sqrt{G(m_0+m_1+m_2)a_2}~~,\label{eq:L2}
\end{eqnarray}
where $\mu_2$ is the reduced mass of the outer orbit $\mu_2=(m_0+m_1)m_2/(m_0+m_1+m_2)$.

Substituting Equations (\ref{eq:APK}) and (\ref{eq:APO}) into (\ref{eq:Gj1})\---(\ref{eq:Ge2}),
the octupole-level secular evolution equations can be obtained.

For the inner orbit, we have
\begin{equation}\label{eq:j1vec}
\begin{split}
\frac{d{\jvec_1}}{dt}&=\frac{3}{4~t_K}\Big[(\jvec_1\cdot\nvec_2)~\jvec_1\times\nvec_2
-5(\evec_1\cdot\nvec_2)~\evec_1\times\nvec_2\Big]\\
&-\frac{75\varepsilon_{\oct}}{64~t_K}\Bigg\{
\bigg[2\Big[(\evec_1\cdot\uvec_2)(\jvec_1\cdot\nvec_2)\\
&+(\evec_1\cdot\nvec_2)(\jvec_1\cdot\uvec_2)\Big]~\jvec_1+2\Big[(\jvec_1\cdot\uvec_2)(\jvec_1\cdot\nvec_2)\\
&-7(\evec_1\cdot\uvec_2)(\evec_1\cdot\nvec_2)\Big]~\evec_1\bigg]\times\nvec_2\\
&+\bigg[2(\evec_1\cdot\nvec_2)(\jvec_1\cdot\nvec_2)~\jvec_1
+\Big[\frac{8}{5}e_1^2-\frac{1}{5}\\
&-7(\evec_1\cdot\nvec_2)^2+(\jvec_1\cdot\nvec_2)^2\Big]~\evec_1\bigg]
\times\uvec_2\Bigg\},
\end{split}
\end{equation}
\begin{equation}\label{eq:e1vec}
\begin{split}
\frac{d{\evec_1}}{dt}&=\frac{3}{4~t_K}\Big[(\jvec_1\cdot\nvec_2)~\evec_1\times\nvec_2
+2~\jvec_1\times\evec_1\\
&-5(\evec_1\cdot\nvec_2)\jvec_1\times\nvec_2\Big]\\
&-\frac{75\varepsilon_{\oct}}{64~t_K}\Bigg\{
\bigg[2(\evec_1\cdot\nvec_2)(\jvec_1\cdot\nvec_2)~\evec_1\\
&+\Big[\frac{8}{5}e_1^2-\frac{1}{5}-7(\evec_1\cdot\nvec_2)^2+(\jvec_1\cdot\nvec_2)^2\Big]~\jvec_1\bigg]\times\uvec_2\\
&+\bigg[2\Big[(\evec_1\cdot\uvec_2)(\jvec_1\cdot\nvec_2)+(\evec_1\cdot\nvec_2)(\jvec_1\cdot\uvec_2)\Big]~\evec_1\\
&+2\Big[(\jvec_1\cdot\nvec_2)(\jvec_1\cdot\uvec_2)-7(\evec_1\cdot\nvec_2)(\evec_1\cdot\uvec_2)\Big]~\jvec_1
\bigg]\times\nvec_2\\
&+\frac{16}{5}(\evec_1\cdot\uvec_2)~\jvec_1\times\evec_1\Bigg\}~~.
\end{split}
\end{equation}

For the outer orbit, we have
\begin{equation}\label{eq:j2vec}
\begin{split}
\frac{d{\jvec_2}}{dt}&=\frac{3}{4t_K}\frac{L_1}{L_2}\Big[(\jvec_1\cdot\nvec_2)~\nvec_2\times\jvec_1
-5(\evec_1\cdot\nvec_2)~\nvec_2\times\evec_1\Big]\\
&-\frac{75\varepsilon_{\oct}}{64t_K}\frac{L_1}{L_2}\Bigg\{
2\Big[(\evec_1\cdot\nvec_2)(\jvec_1\cdot\uvec_2)~\nvec_2\\
&+(\evec_1\cdot\uvec_2)(\jvec_1\cdot\nvec_2)~\nvec_2+(\evec_1\cdot\nvec_2)(\jvec_1\cdot\nvec_2)~\uvec_2\Big]\times\jvec_1\\
&+\bigg[2(\jvec_1\cdot\uvec_2)(\jvec_1\cdot\nvec_2)~\nvec_2-
14(\evec_1\cdot\uvec_2)(\evec_1\cdot\nvec_2)~\nvec_2\\
&+\Big[\frac{8}{5}e_1^2-\frac{1}{5}-7(\evec_1\cdot\nvec_2)^2+(\jvec_1\cdot\nvec_2)^2\Big]~\uvec_2\frac{}{}\bigg]\times\evec_1
\Bigg\},\\
\end{split}
\end{equation}

\begin{equation}\label{eq:e2vec}
\begin{split}
\frac{d{\evec_2}}{dt}&=\frac{3}{4t_K\sqrt{1-e_2^2}}\frac{L_1}{L_2}\bigg[
(\jvec_1\cdot\nvec_2)~\evec_2\times\jvec_1-5(\evec_1\cdot\nvec_2)\evec_2\times\evec_1\\
&-\Big[\frac{1}{2}-3e_1^2+\frac{25}{2}(\evec_1\cdot\nvec_2)^2-\frac{5}{2}(\jvec_1\cdot\nvec_2)^2\Big]\nvec_2\times\evec_2\bigg]\\
&-\frac{75}{64t_K}\frac{\varepsilon_{\oct}}{\sqrt{1-e_2^2}}\frac{L_1}{L_2}
\Bigg\{2\Big[(\evec_1\cdot\nvec_2)(\jvec_1\cdot\evec_2)~\uvec_2\\
&+(\jvec_1\cdot\nvec_2)(\evec_1\cdot\evec_2)~\uvec_2+\frac{1-e_2^2}{e_2}(\evec_1\cdot\nvec_2)
(\jvec_1\cdot\nvec_2)~\nvec_2\Big]\times\jvec_1\\
&+\bigg[2(\jvec_1\cdot\evec_2)(\jvec_1\cdot\nvec_2)~\uvec_2
-14(\evec_1\cdot\evec_2)(\evec_1\cdot\nvec_2)~\uvec_2\\
&+\frac{1-e_2^2}{e_2}\Big[\frac{8}{5}e_1^2-\frac{1}{5}-7(\evec_1\cdot\nvec_2)^2+(\jvec_1\cdot\nvec_2)^2\Big]~\nvec_2\bigg]
\times\evec_1\\
&-\bigg[2\left(\frac{1}{5}-\frac{8}{5}e_1^2\right)(\evec_1\cdot\uvec_2)~\evec_2\\
&+14(\evec_1\cdot\nvec_2)(\jvec_1\cdot\uvec_2)(\jvec_1\cdot\nvec_2)~\evec_2
+7(\evec_1\cdot\uvec_2)\Big[\frac{8}{5}e_1^2\\
&-\frac{1}{5}-7(\evec_1\cdot\nvec_2)^2+(\jvec_1\cdot\nvec_2)^2\Big]~\evec_2
\bigg]\times\nvec_2\Bigg\}~~.
\end{split}
\end{equation}

In the above, we have defined the (quadrupole) Kozai timescale as
\begin{equation}
t_K\equiv\frac{L_1}{\mu_1\Phi_0}=\frac{1}{n_1}\left(\frac{m_0+m_1}{m_2}\right)\left(\frac{a_2}{a_1}\right)^3(1-e_2^2)^{3/2}~,
\end{equation}
where $n_1\equiv\sqrt{G(m_0+m_1)/a_1^3}$ is the mean motion of the inner binary.
Equations (\ref{eq:j1vec})\---(\ref{eq:e2vec}) describe the long-term evolution of the inner and outer binaries
for all mass ratios.
Our equations are equivalent to those presented in \citet{Petrovich},
although they are in a somewhat different form.

Often times, the triple system contains a body of
much smaller mass than the other two, such as in the case of a planet around
one member of a binary. In this case, the planet can be considered
to be a particle of effective zero mass to a very good approximation.
In this ``test-particle limit" \citep[e.g.][]{LS, Katz PRL, Smadar 2013b},
the outer orbit contains the totality of the angular momentum in the system,
and consequently remains fixed in time. To derive
the test-particle limit from Equations~(\ref{eq:j1vec})\---(\ref{eq:e2vec}), we
take the limit $L_1/L_2\rightarrow 0$, for which we confirm that
$d\jvec_2/dt=d\evec_2/dt=0$. In this case, the triad $( \hat{\mathbf{x}}, \hat{\mathbf{y}}, \hat{\mathbf{z}})$ 
is fixed in space, thus, for the sake of clarity, we relabel these vectors with lab-frame coordinates
$( \hat{\mathbf{x}}, \hat{\mathbf{y}}, \hat{\mathbf{z}})$ in Equations~(\ref{eq:j1vec})\---(\ref{eq:e1vec})
and in Equations~(\ref{eq:APK}) and~(\ref{eq:APO}). With these replacements,
we recover the expressions of \citet{Katz PRL} for the potentials  and for
$d\jvec_1/dt$ and $\evec_1/dt$ for a test particle.

%%%%%%%%%%%%%%%%%%%%%%%%%%%%%
\subsection{Equations of motion in orbital elements form}
The secular equations for the orbital elements
of the inner and outer orbits have been the focus of
previous work on hierarchical triple systems  \citep{Ford, Smadar 2011}.
With the introduction of Delaunay variables, conjugate pairs of coordinates
and momenta can be defined,
the equations of motion of the orbital elements
can be directly derived using Hamilton's equations \citep[e.g.][]{MD}.

Instead of using Hamilton equations, one can convert the vector equations  (Equations~\ref{eq:j1vec}\---\ref{eq:e2vec})
into the orbital element form by expressing the Cartesian components of the vectors $\jvec_\alpha$, $\evec_\alpha$
(with $\alpha=1,2$) in terms of the orbital eccentricity  $e_\alpha$, inclination $i_\alpha$,
argument of periapse $\omega_\alpha$ and  longitude of ascending nodes $\Omega_\alpha$
\citep[e.g.][]{MD}:
\begin{eqnarray}
&&\jvec_\alpha =\sqrt{1-e_\alpha^2}
\begin{pmatrix}
\sin i_\alpha \sin\Omega_\alpha \\
-\sin i_\alpha \cos\Omega_\alpha \\
\cos i_\alpha
\end{pmatrix},\label{eq:proj}\\
&&\evec_\alpha =e_\alpha
\begin{pmatrix}
\cos\omega_\alpha\cos\Omega_\alpha-\sin\omega_\alpha\cos i_\alpha\sin\Omega_\alpha \\
\cos\omega_\alpha\sin\Omega_\alpha+\sin\omega_\alpha\cos i_\alpha\cos\Omega_\alpha \\
\sin\omega_\alpha\sin i_\alpha~.
\end{pmatrix},\label{eq:proe}
\end{eqnarray}
Here the angles are defined respect to a fixed coordinate frame
in which $z$-axis is aligned with the (conserved) total angular momentum,
while the $x$-$y$ plane coincides with the so-called invariable plane.
Thus the relative inclination between the two orbits is  $\I\equiv i_1+i_2$.
Because of angular momentum conservation,
the condition $\Delta\Omega=\Omega_1-\Omega_2=\pi$ is satisfied.

By substituting Equations~(\ref{eq:proj})~and~(\ref{eq:proe}) into Equations~(\ref{eq:j1vec})\---(\ref{eq:e2vec}),
we can solve for $de_\alpha/dt$, $di_\alpha/dt$, $d\omega_\alpha/dt$ and
$d\Omega_\alpha/dt$ (for $\alpha=1,2$).
Since $d\Omega_1/dt=d\Omega_2/dt$,
the system is determined by seven independent differential equations.
These equations are listed in Appendix~\ref{sec:Full} (see Equations~\ref{eq:fulle1}\---\ref{eq:fullcos}).
Although different
in form, we have checked that these equations are equivalent to those
presented by \citet{Smadar 2013b}.

For the test mass case ($m_1\ll m_0$), we take the limit $L_1/L_2\rightarrow0$
in Equations~(\ref{eq:fulle1})\---(\ref{eq:fullcos}).
Also, note that in this limit, $\varepsilon_{\oct}\rightarrow (a_1/a_2)e_2/(1-e_2^2)$, $i_1 \rightarrow\I $, $i_2\rightarrow 0$.
We choose $\uvec_2= \hat{\mathbf{x}}$, $\hat{\mathbf{v}}_2=\hat{\mathbf{y}}$,
$\nvec_2= \hat{\mathbf{z}}$ and $\varpi_2=0$,
where $\omega_2=\varpi_2-\Omega_2=\varpi_2-\Omega_1+\pi$.
Accordingly, the secular evolution equations become
\begin{equation}\label{eq:testPe}
\begin{split}
\frac{de_1}{d\tau}&=\frac{15}{8}e_1\sqrt{1-\e1}\sin^2\!i_1\sin2\omega_1\\
&-\frac{15\sqrt{1-\e1}\varepsilon_{\oct}}{512}\bigg\{\cos\Omega_1\Big[(4+3\e1)(3+5\cos2i_1)\\
&\times\sin\omega_1+210~\e1\sin^2\!i_1\sin3\omega_1\Big]+2\cos i_1\\
&\times\cos\omega_1\sin\Omega_1\Big[15(2+5\e1)\cos2i_1\\
&+7(30\e1\cos2\omega_1\sin^2\!i_1-2-9\e1)\Big]\bigg\},
\end{split}
\end{equation}
\begin{equation}\label{eq:testPi}
\begin{split}
\frac{di_1}{d\tau}&=-\frac{15}{16}\frac{\e1\sin2i_1\sin2\omega_1}{\sqrt{1-\e1}}
+\frac{15~e_1\varepsilon_{\oct}}{256\sqrt{1-\e1}}\\
&\times\bigg\{10\sin2i_1\cos\Omega_1\sin\omega_1(2+5\e1+7\e1\cos2\omega_1)\\
&-\cos\omega_1\sin i_1\sin\Omega_1\Big[26+37\e1-35\e1\cos2\omega_1\\
&-15\cos2i_1(7\e1\cos2\omega_1-2-5\e1)\Big]\bigg\},
\end{split}
\end{equation}
\begin{equation}\label{eq:testPO}
\begin{split}
\frac{d\Omega_1}{d\tau}&=\frac{3}{4}\frac{\cos i_1(5\e1\cos^2\!\omega_1-4\e1-1)}{\sqrt{1-\e1}}\\
&+\frac{15~e_1\varepsilon_{\oct}}{128\sqrt{1-\e1}}\bigg\{20\cos i_1\cos\omega_1(2+5\e1-7\e1\cos2\omega_1)\\
&\times\cos\Omega_1+\Big[35\e1(1+3\cos2i_1)\cos2\omega_1-46\\
&-17\e1-15(6+\e1)\cos2i_1\Big]\sin\omega_1\sin\Omega_1\bigg\},
\end{split}
\end{equation}
\begin{equation}\label{eq:testPo}
\begin{split}
\frac{d\omega_1}{d\tau}&=\frac{3}{4}\frac{2(1-\e1)+5\sin^2\!\omega_1(\e1-\sin^2\!i_1)}{\sqrt{1-\e1}}\\
&+\frac{15\varepsilon_{\oct}}{64}
\Bigg\{\frac{e_1\cos i_1}{\sqrt{1-\e1}}\bigg[\sin\omega_1\sin\Omega_1\\
&\times\Big[10(3\cos^2\!i_1-1)(1-\e1)+A\Big]-5B\cos i_1\cos\Theta\bigg]\\
&-\frac{\sqrt{1-\e1}}{e_1}\Big[10\sin\omega_1\sin\Omega_1\cos i_1\\
&\times\sin^2\!i_1(1-3\e1)+\cos\Theta(3A-10\cos^2\!i_1+2)\Big]\Bigg\},
\end{split}
\end{equation}
where we have introduced the dimensionless time $\tau\equiv t/t_K$
and
\begin{eqnarray}
~~~~~~&&A\equiv4+3\e1-\frac{5}{2}B\sin^2\!i_1~, ~~\\
&&B\equiv2+5\e1-7\e1\cos2\omega_1~,\\
&&\cos\Theta\equiv\cos\omega_1\cos\Omega_1-\cos i_1\sin\omega_1\sin\Omega_1~.\label{eq:testPcos}
\end{eqnarray}

When the octupole terms are ignored (i.e., $\varepsilon_\oct=0$),
Equations~(\ref{eq:testPe})--(\ref{eq:testPo}) reduce to the orbital
element equations of motion found in \citet[][Eqs. 5]{Innanen} for Lidov-Kozai
oscillations%%%%%%%%%%%%%%%%%%%%%%%%%
\footnote{Note that, due to a formatting error, the evolution
equation for $\omega_1$ in \citet{Innanen} appears with the term $\sqrt{1-e_1^2}$ in the
numerator instead of the denominator.%%%%%%%%%%%%%%%%%%%%%%%
}.

%%%%%%%%%%%%%%%%%%%%%%%%%%%%%%%%%%%%%%%%%%%%%%%
%%%%%%%%%%%%%%%%%%%%%%%%%%%%%%%%%%%%%%%%%%%%%%%
\begin{figure}
\begin{centering}
\includegraphics[width=8.3cm]{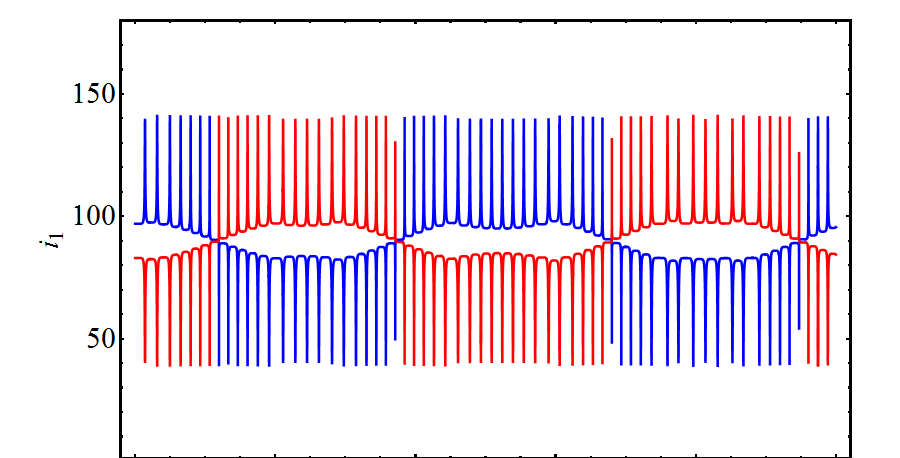}
\includegraphics[width=8.3cm]{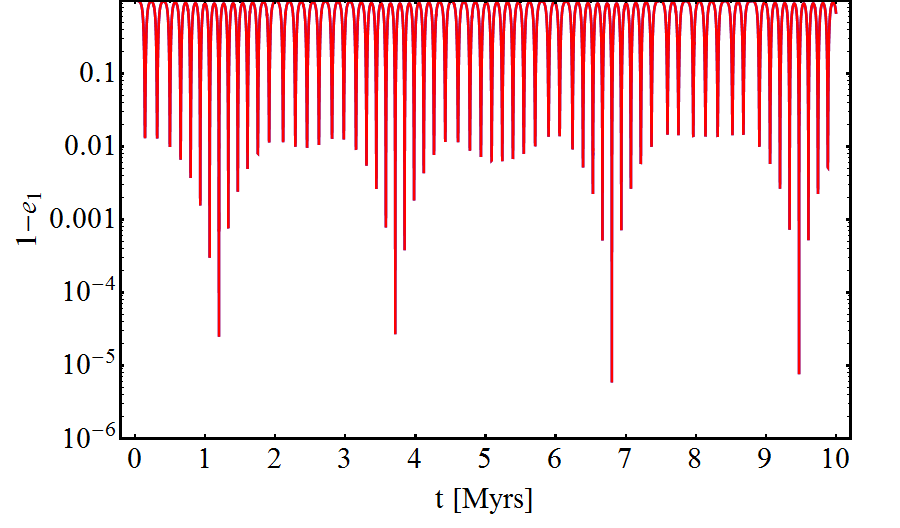}
\caption{
Evolution of inclination $i_1$ and eccentricity $e_1$ 
from numerical integration of 
Equations (\ref{eq:testPe})\---(\ref{eq:testPo})
for a prograde (red) and a retrograde (blue) inner binary.
The system parameters are $m_0=1M_\odot$, $m_1\ll m_0$, 
$m_2=1M_\odot$, $a_1=1\au$, $a_2=100\au$ and
$e_2=0.8$. We set $e_1=10^{-3}$ initially. Inclinations are initialized 
at $\pm7^\circ$ from $90^\circ$.
Red lines:~$\I=i_1=83^\circ$, $\omega=0^\circ$, $\Omega=180^\circ$;
Blue lines:~$\I=i_1=97^\circ$, $\omega=180^\circ$, $\Omega=0^\circ$.
In the lower panel, the blue and red curves exactly overlap.
}
\label{fig:STP}
\end{centering}
\end{figure}
%%%%%%%%%%%%%%%%%%%%%%%%%%%%%%%%%%%%%%%%%%%%%%%
%%%%%%%%%%%%%%%%%%%%%%%%%%%%%%%%%%%%%%%%%%%%%%%

%%%%%%%%%%%%%%%%%%%%%%%%%%%%%%%%%%%%%%%%%%%%%%%
%%%%%%%%%%%%%%%%%%%%%%%%%%%%%%%%%%%%%%%%%%%%%%%
\begin{figure}
\begin{centering}
%\vskip -0.5truecm
\includegraphics[width=8.3cm]{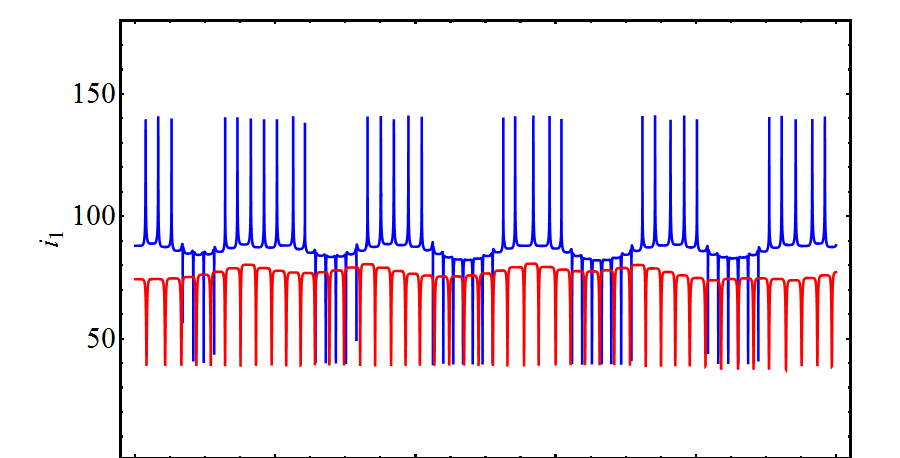}
\includegraphics[width=8.3cm]{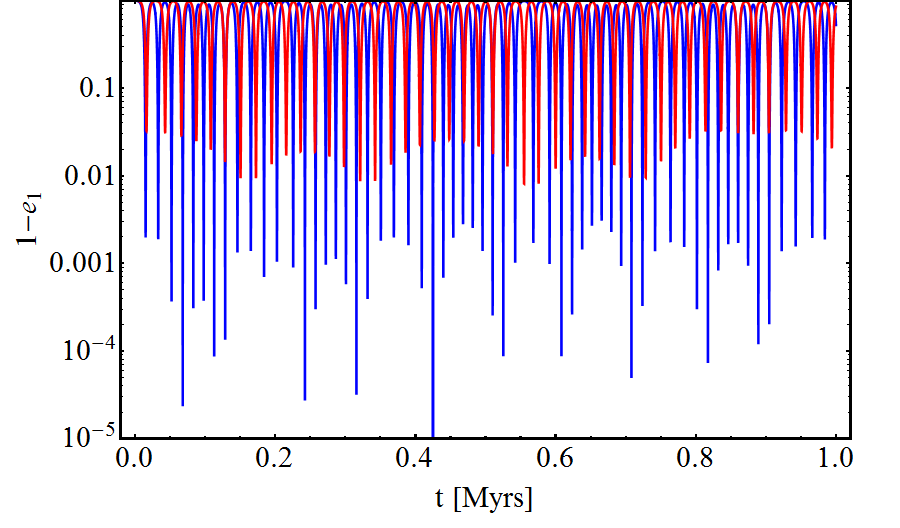}
%\vskip -3.truecm
\caption{
As in Figure~\ref{fig:STP}, evolution of inclination $i_1$ and eccentricity $e_1$
from numerical integration of Equations (\ref{eq:fulle1})\---(\ref{eq:fullcos}) for
prograde (red) and retrograde (blue) inner binaries, this time for a general triple with comparable masses.
The system parameters are $m_0=1M_\odot$, $m_1=0.5M_\odot$, $m_2=1M_\odot$, $a_1=10\au$
and $a_2=100\au$. Eccentricities are initialized as  $e_1=10^{-3}$ and $e_2=0.5$.
Red lines:~$\I=83^\circ$, $i_1=74.3^\circ$, $i_2=8.7^\circ$, $\omega_1=0^\circ$, $\Omega_1=0^\circ$;
Blue lines:~$\I=97^\circ$, $i_1=88^\circ$, $i_2=9^\circ$, $\omega_1=180^\circ$, $\Omega_1=180^\circ$.
}
\label{fig:SCM}
\end{centering}
\end{figure}
%%%%%%%%%%%%%%%%%%%%%%%%%%%%%%%%%%%%%%%%%%%%%%%
%%%%%%%%%%%%%%%%%%%%%%%%%%%%%%%%%%%%%%%%%%%%%%%

%%%%%%%%%%%%%%%%%%%%%%%%
\subsection{Reflection symmetry of the equations of motion}\label{sec:symmetry}
In the test-particle limit ($m_1\ll m_0$), the vector equations (\ref{eq:j1vec})\---(\ref{eq:e1vec})
are symmetric under reflections of the inner binary. If we perform
the replacement $\jvec_1\rightarrow-\jvec_1$ (leaving $\evec_1$ unchanged),
we find that $d\jvec_1/dt\rightarrow-d\jvec_1/dt$ and $d\evec_1/dt\rightarrow-d\evec_1/dt$,
which may be interpreted as reversing the direction of time.  In terms of orbital elements,
this reflection operation is equivalent to changing 
$i_{1}\rightarrow\pi-i_{1}$, $\omega_{1}\rightarrow\pi-\omega_{1}$
and $\Omega_{1}\rightarrow\Omega_{1}+\pi$. Performing this replacement
in Equations~(\ref{eq:testPe})--(\ref{eq:testPo}), we obtain
$de_1/d\tau\rightarrow-de_1/d\tau$, $di_1/d\tau\rightarrow-di_1/d\tau$,
$d\Omega_1/d\tau\rightarrow-d\Omega_1/d\tau$ and 
$d\omega_1/d\tau\rightarrow-d\omega_1/d\tau$, as expected.

Figure~\ref{fig:STP}
shows the integration of Equations (\ref{eq:testPe})\---(\ref{eq:testPo})
for two configurations differing solely on the orientation of the
$\jvec_1$ vector (this reflection is carried out by changing
the initial conditions
$i_{1,0}\rightarrow\pi-i_{1,0}$, $\omega_{1,0}\rightarrow\pi-\omega_{1,0}$
and $\Omega_{1,0}\rightarrow\Omega_{1,0}+\pi$). The evolution of eccentricity is
indistinguishable between the prograde (red curves) and retrograde (blue curves)
cases, while the inclination angle $i_1$ shows a reflection symmetry around $90^\circ$,
evolving in an identical manner in both cases except for a phase offset
of half of what can be interpreted as an ``octupole period" \citep{Teyssandier}.

For the comparable-mass case ($m_1\sim m_0$),
the inner and outer binaries evolve together, exchanging angular momentum.
Under reflection operation ($\jvec_1\rightarrow-\jvec_1$),
we find that $d\jvec_1/dt\rightarrow-d\jvec_1/dt$, $d\evec_1/dt\rightarrow-d\evec_1/dt$,
$d\jvec_2/dt\rightarrow d\jvec_2/dt$ and $d\evec_2/dt\rightarrow d\evec_2/dt$,
i.e., the symmetry of the equations is broken.
Figure \ref{fig:SCM} shows the numerical integration in the case of general masses,
for two configurations with similar reflection operation as in Figure \ref{fig:STP}.
In this case, it is apparent that there is no reflection symmetry
between the prograde (red curves) and retrograde (blue curves) initial conditions.
Even the ``octupole periods" are different.

%%%%%%%%%%%%%%%%%%%%%%%%%%%%%%%%%%%%%%%%%%%%%%%%%%%%%%%%%%%
\section{Effects of short-range forces}
The high eccentricity phase of a Lidov\---Kozai cycle can be severely modified
if the inner binary separation at pericenter is sufficiently small for additional
forces to overcome the tidal torque exerted by the outer binary. If the energy
associated to these extra forces $\Phi_\mathrm{extra}$ surpasses the interaction potential $\Phi$
of Equation~(\ref{eq:Ham}), the Lidov\---Kozai mechanism is said to be
``arrested" \citep[e.g.][]{WM}.

Here we study the effects of short-range forces on the eccentricity evolution of
triple system consisting of a Jupiter-mass planet orbiting the primary star
of a binary.  The short range effects we consider include (1) precession of
periapse due to GR, (2) tidal bulge of the planet induced by the star,
and (3) planet oblateness due to rotation.

%%%%%%%%%%%%%%%%%%%%%%%%%%%%%%%%%%%%
\subsection{Conservative short-range forces}
In the absence of energy dissipation (e.g., tidal friction or gravitational wave radiation),
the energy of the system $\mathcal{H}=\mathcal{H}_1+\mathcal{H}_2+\Phi+\Phi_\mathrm{extra}$
is conserved, and so is its orbit-averaged version. Since
the semimajor axes of the inner and outer orbits are constant,
we only need to consider the conserved potential 
\begin{equation}\label{eq:PT}
\langle\Phi_{\mathrm{tot}}\rangle=\langle\Phi\rangle+\langle\Phi_{\gr}\rangle
+\langle\Phi_{\tide}\rangle+\langle\Phi_{\rot}\rangle~~,
\end{equation}
where, as before, $\Phi=\Phi_\K+\Phi_\oct$.

The post-Newtonian potential associated with periastron advance is \citep[e.g.,][]{Eggleton}
\begin{equation}\label{eq:PGR}
\begin{split}
\langle\Phi_{\gr}\rangle&=-\frac{3G^2m_0m_1(m_0+m_1)}{a_1^2c^2}\frac{1}{(1-\e1)^{1/2}}\\
&=-\varepsilon_{\gr}\mu_1\Phi_0\frac{1}{(1-\e1)^{1/2}}~,
\end{split}
\end{equation}
where we have defined the dimensionless parameter
\begin{equation}\label{eq:R11}
\varepsilon_{\gr}\equiv\frac{3G(m_0+m_1)^2a_2^3(1-e_2^2)^{3/2}}{a_1^4c^2m_2}~.
\end{equation}

The potential due to the non-dissipative tidal bulge on $m_1$ is
\begin{equation}\label{eq:PTide}
\begin{split}
\langle\Phi_{\tide}\rangle&=-\frac{G}{a_1^6}\frac{1+3\e1+\frac{3}{8}e_1^4}{(1-\e1)^{9/2}}\left(\frac{}{}
m_0^2k_{2,1}R_1^5\right)\\
&=-\varepsilon_{\tide}\frac{\mu_1\Phi_0}{15}\frac{1+3\e1+\frac{3}{8}e_1^4}{(1-\e1)^{9/2}}~,
\end{split}
\end{equation}
where
\begin{equation}\label{eq:R12}
\varepsilon_{\tide}\equiv\frac{15m_0(m_0+m_1)a_2^3(1-e_2^2)^{3/2}k_{2,1}R_1^5}{a_1^8m_1m_2}~,\\
\end{equation}
and $k_{2,1}$, $R_1$ are the tidal Love number and the radius of $m_1$, respectively.
The potential energy associated with the rotation-induced oblateness of $m_1$ is
\begin{equation}\label{eq:PRot0}
\langle\Phi_{\rot}\rangle=-\frac{G m_0 (I_3-I_1)_1}{2 a_1^3(1-\e1)^{3/2}}~.
\end{equation}
Here,
\begin{equation}
(I_3-I_1)_1=\frac{2}{3}k_{q,1}m_1 R_1^2\left(\frac{\Omega_{1s}^2}{{G m_1}/{R_1^3}}\right)~,
\end{equation}
where $k_{q,1}$ is the apsidal motion constant and $\Omega_{1s}$ is the spin rate of $m_1$,
and we have assumed that the spin vector of $m_1$ is aligned with the angular momentum vector $\jvec_1$ of
the inner orbit. We can rewrite Equation~(\ref{eq:PRot0}) as
\begin{equation}\label{eq:PRot}
\langle\Phi_{\rot}\rangle=-\varepsilon_{\rot}\frac{\mu_1\Phi_0}{3}\frac{1}{(1-\e1)^{3/2}}~,
\end{equation}
where
\begin{equation}\label{eq:R13}
\varepsilon_{\rot}\equiv\frac{(m_0+m_1)a_2^3(1-e_2^2)^{3/2}k_{q,1}\Omega_{1s}^2R_1^5}{Ga_1^5m_1m_2}~.
\end{equation}

The three dimensionless parameters $\varepsilon_{\gr}$, $\varepsilon_{\tide}$ and $\varepsilon_{\rot}$ quantify the relative importance
of the short-range potential terms respect to the quadrupole potential $\Phi_\K$.

Since the three short-range potentials (Equations~\ref{eq:PGR},~\ref{eq:PTide}~and~\ref{eq:PRot})
depend on the orbital vectors solely through
$e_1=|\evec_1|$, only the evolution equations for $\evec_1$ is modified
(see Equations~\ref{eq:Gj1} and~\ref{eq:Ge1}).
 These extra
forces induce an additional precession of $\evec_1$ around $\jvec_1$:
\begin{equation}\label{eq:extrae1}
\left.\frac{d\evec_1}{dt}\right|_{\extra}=\Dot \omega_\extra~\nvec_1\times\evec_1~,
\end{equation}
where \citep[e.g.,][]{FT}.
\begin{equation}\label{eq:extraomega}
\Dot \omega_{\extra}=-\frac{\sqrt{1-\e1}}{e_1L_1}\frac{\partial\langle \Phi_{\extra}\rangle}{\partial e_1}~.
\end{equation}
Thus, the GR-induced precession rate is:
\begin{equation}\label{eq:omegagr}
\Dot\omega_\gr=\frac{\varepsilon_{\gr}}{t_K}\frac{1}{1-\e1}~.
\end{equation}
Similarly, for the static tide, we have
\begin{equation}\label{eq:omegatide}
\Dot\omega_\tide=\frac{\varepsilon_{\tide}}{t_K}\frac{1+\frac{3}{2}\e1+\frac{1}{8}e_1^4}{(1-\e1)^5}~,
\end{equation}
and for the rotation-induced planet oblateness
\begin{equation}\label{eq:omegarot}
\Dot\omega_\rot=\frac{\varepsilon_{\rot}}{t_K}\frac{1}{(1-\e1)^2}~.
\end{equation}
To obtain an estimate of the relative importance of $\Dot \omega_\tide$ and $\Dot \omega_\rot$,
we may consider a pseudo-synchronized planet spin, that is,
the planet rotation rate $\Omega_{1s}$ is of order the orbital frequency of periapse $n_1(1-e_1^2)^{-3/2}$.
In the weak friction theory of equilibrium tides, the pseudo-synchronized rotation rate is
given by \citep[e.g.,][]{Alexander,Hut}
\begin{equation}
\left(\frac{\Omega_{1s}}{n_1}\right)_{\mathrm{ps}}\equiv f_{\mathrm{ps}}(e_1)=
\frac{1+\frac{15}{2}\e1+\frac{45}{8}e_1^4+\frac{5}{16}e_1^6}{1+3\e1+\frac{3}{8}e_1^4}
\frac{1}{(1-\e1)^{3/2}}~.
\end{equation}
Thus, in Equation~(\ref{eq:omegarot}), we have $\varepsilon_\rot=\varepsilon_\rot'f_{\mathrm{ps}}^2(e_1)$
with
\begin{equation}
\varepsilon_\rot'=\frac{(m_0+m_1)^2 a_2^3 (1-e_2^2)^{3/2}k_{q,1}R_1^5}{a_1^8m_1m_2}~.
\end{equation}
Comparing with $\Dot \omega_\tide$, we have $\Dot \omega_\tide/\Dot \omega_\rot\simeq(15k_{2,1}/k_{q,1})$.
Since $k_{2,1}\simeq 2k_{q,1}$,
we find $\Dot \omega_\tide\gg\Dot \omega_\rot$ for synchronized rotation. Thus, in the vast majority
of our examples, the effect of tides will dominate over the effect of the rotational bulge.

%%%%%%%%%%%%%%%%%%%%%%%%%%%%%%%%%%%%
\subsection{Numerical integrations}\label{sec:examples}
Before systematically examining the parameter space in $\varepsilon_\oct$,
$\varepsilon_\mathrm{GR}$, $\varepsilon_\mathrm{Tide}$ and $\varepsilon_\mathrm{Rot}$,
we consider a few examples to illustrate how SRFs affect Lidov\---Kozai oscillations.

%%%%%%%%%%%%%%%%%%%%%%%%%%%%%%%%%%%%%%%%%
%%%%%%%%%%%%%%%%%%%%%%%%%%%%%%%%%%%%%%%%%
\begin{figure}
\begin{centering}
\includegraphics[width=8.2cm]{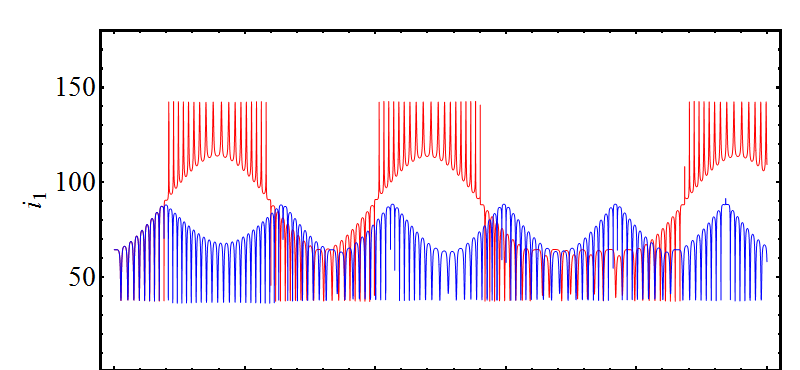}
\includegraphics[width=8.2cm]{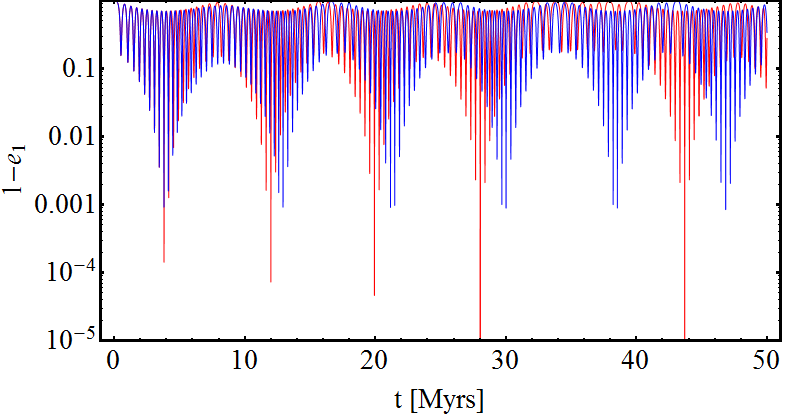}
\caption{Numerical integration of a system with parameters $m_0=1M_\odot$, $m_1=1M_J$,  $m_2=40M_J$ $a_1=6\au$
and $a_2=100\au$.
Initial conditions are $e_1=0.001$, $e_2=0.6$, $i_1=64.7^{\circ}$, $i_2=0.3^{\circ}$, $\I=65^{\circ}$, $\omega_1=45^{\circ}$, $\omega_2=0^{\circ}$, {$\Omega_1=0^\circ$ and $\Omega_2=180^\circ$.}
Total integration time is 50~Myr$\sim360t_K$. The red lines are from the integration of the pure Lidov\---Kozai effect to octupole order, while the blue lines are the results of integration
including SRFs. Orbital flips seem to be entirely suppressed. However, when extending the integration time to $\sim300$ Myrs~$\sim2100t_K$, the inner orbit eventually 
flips to retrograde in the interval of 90--130 Myr before going back to its original orientation.
}
\label{fig:case91}
\end{centering}
\end{figure}
%%%%%%%%%%%%%%%%%%%%%%%%%%%%%%%%%%%%%%%%%
%%%%%%%%%%%%%%%%%%%%%%%%%%%%%%%%%%%%%%%%%

In Figure~\ref{fig:case91}, we show the evolution of a triple system with $\varepsilon_\oct=0.056$
and initial mutual inclination $i_\mathrm{tot}=65^\circ$
obtained by numerical integration of the equations of motion in orbital elements form (see Appendix~\ref{sec:Full}).
When SRFs are
ignored (red curves), the inner orbit evolves into a highly eccentric state over a timescale of order
 $\varepsilon_\oct^{-1} t_K$,
reaching values as extreme as $1-e_1<10^{-5}$ (red curves, bottom panel). In this example,
eccentricity maxima of $e_1\rightarrow1$ are always accompanied by orbital flips (red curves, top panel),
i.e., the $z$-component of $\jvec$ reverses its sign. Orbital flips are always tied
to eccentricity maxima and correspond to $\jvec_1$ shrinking going through the origin ($|\jvec_1|=0$)
as $e_1\rightarrow1$ \citep[e.g.,][]{Katz PRL}. In some cases, it is possible to
even derive a closed-form solution of this behaviour over long timescales
provided the slowly varying quantity $\tfrac{4}{3}\Phi_\K/(\mu_1\Phi_0)-\tfrac{1}{2}(\jvec_1\cdot\hat{\mathbf{z}})^2+\tfrac{1}{6}$
remains positive at all times \citep[e.g.,][]{Katz PRL}. Assuming that
the maximum eccentricity of a Lidov\---Kozai cycle is reached when $j_z\equiv\jvec_1\cdot\hat{\mathbf{z}}$ crosses zero,
\citet{Katz PRL} find that the maximum $e_1$ scales with $\varepsilon_\oct$ roughly as $\sim\sqrt{1-\varepsilon_\oct^2}$.
If this theoretical maximum cannot be reached owing to additional effects such as SRFs, then we
expect that $j_z$ will be unable to come arbitrarily close to zero, and therefore orbital flips will not be allowed.

Indeed, when SRFs are included (blue curves), the maximum eccentricity
is capped down to values such that $1-e_1\simeq10^{-3}$
(blue curves, bottom panel). Although still large, this upper limit to the eccentricity is sufficient
to introduce a lower limit to $|\mathbf{j}_1|$ (see Section~\ref{sec:max_ecc} below)
such that $j_z$ cannot reverse signs under the criterion introduced by
\citet{Katz PRL}. As a result, we see no orbital flips in this example (red curves, top panel).

\begin{figure}
\begin{centering}
\includegraphics[width=8.2cm]{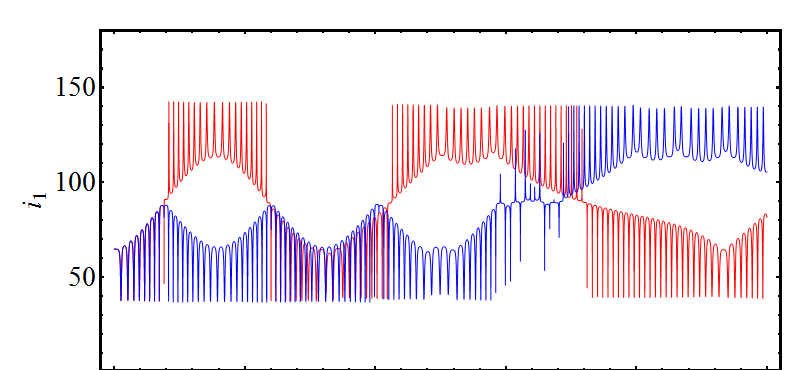}
\includegraphics[width=8.2cm]{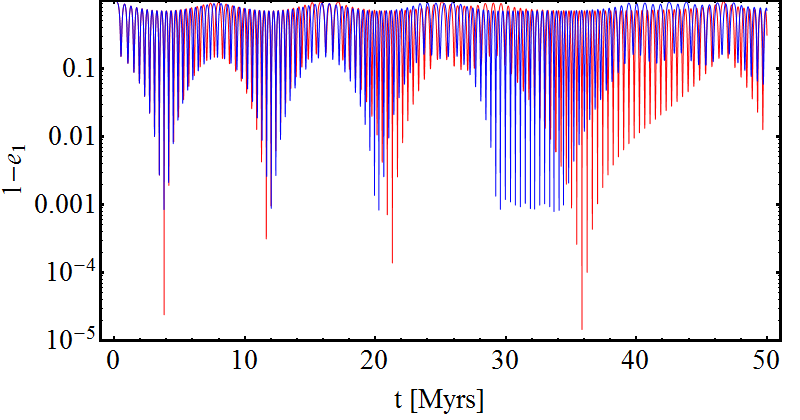}
\caption{{Same initial conditions as} in Figure~\ref{fig:case91}, but slightly changing the initial inclination:  $\I=65.3783^{\circ}$ with $i_1=64.9734^{\circ}$ and $i_2=0.4049^{\circ}$.
}
\label{fig:case92}
\end{centering}
\end{figure}

Surprisingly, however, orbital flips are not prohibited in every case.
In Figure~\ref{fig:case92}, we present an example of a system which exhibits an orbital
flip even though the maximum allowed eccentricity has been reduced by SRFs.
As in the previous example, we integrate a triple system with $\varepsilon_\oct=0.056$ with and without SRFs
(blue and red curves respectively). This time, the initial conditions are modified slightly,
changing the initial mutual inclination angle from $i_\mathrm{tot}=65^\circ$ to
$\I=65.3783^{\circ}$. Over the first half of the integration, the evolution of eccentricity (bottom panel)
and inclination (top panel) in Figure~\ref{fig:case92} closely resemble of those of Figure~\ref{fig:case91}.
However, after 25 Myrs, the two systems start following entirely differently trajectories, despite the very small difference
in initial conditions. In addition, the figure shows that this system finds a way to
cause an orbital flip (i.e. $j_z$ crosses zero), despite that the eccentricity is not allowed to exceed $e_1=1-10^{-3}$
just as in Figure~\ref{fig:case91}. From Figure~\ref{fig:case92}, we conclude that (1) orbital flips can still take place
in presence of SFRs that are strong enough to limit the eccentricity maximum, and (2) that the inclination
of the inner binary may exhibit chaotic behaviour \citep[e.g.][]{Li 2014}, and that conservative SRFs modify
do not necessarily suppress this erratic evolution. To check whether the example of Figure~\ref{fig:case91}
has entirely suppressed orbital flips or if it is just a matter of time before it finds a channel to cross $j_z=0$, we integrate
the system for $300$~Myr~$\sim2100t_K$. Indeed, we confirm that after a very long time $\sim500t_K$, this system also undergoes
an orbital flip that lasts for $300t_K$ before returning to its original orientation.

\begin{figure}
\begin{centering}
\includegraphics[width=8.cm]{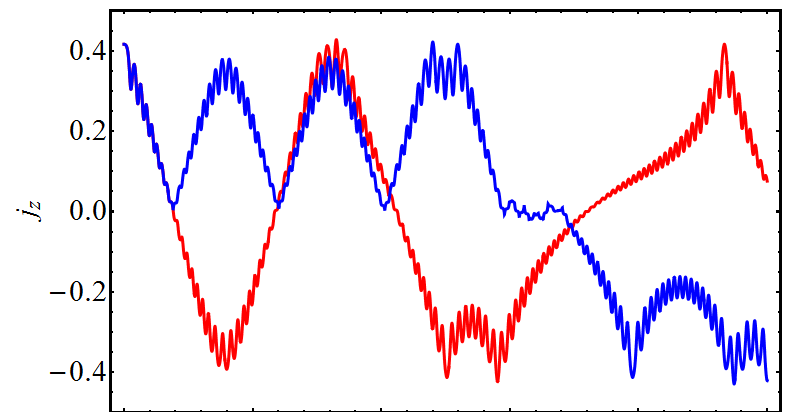}
\includegraphics[width=8.cm]{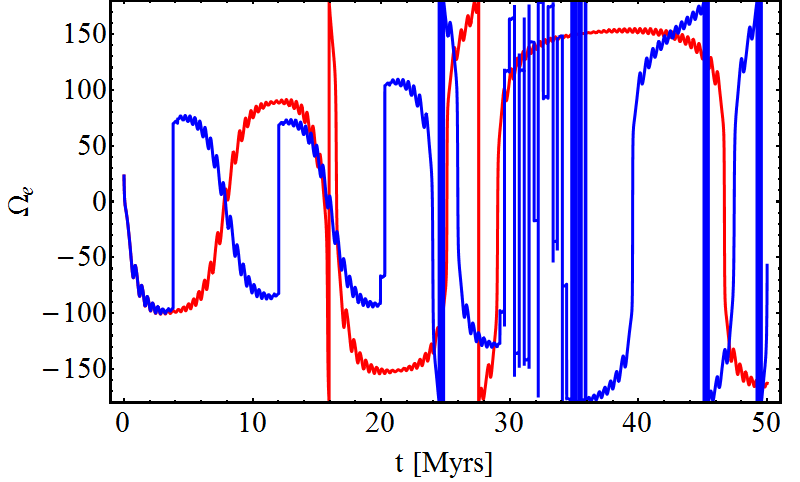}
\caption{Evolution of $j_z=\jvec_1\cdot\hat{\mathbf{z}}$ and  $\Omega_e\equiv \arctan(e_y/e_x)$ for the case 
corresponding to Figure \ref{fig:case92}. The red lines
are from the integration of pure Lidov\---Kozai effect in octupole order, while the blue lines are the results of integration
including SRFs. When SRFs limit the maximum eccentricity, flips are suppressed if $\Omega_e$ is bounded, but they become 
possible once again if $\Omega_e$ is circulating.
}
\label{fig:case93}
\end{centering}
\end{figure}

Figure~\ref{fig:case93} shows the evolution of $j_z$ and the angle $\Omega_e\equiv \arctan(e_y/e_x)$ \citep[see][]{Katz PRL}
for the example of Figure~\ref{fig:case92}. The red curves show the evolution of the system in the absence of SRFs,
where a regular oscillation of $j_z$ at earlier times transitions to a different regime after $\sim35$~Myrs.  When
SRFs are included (blue curves), the evolution of $j_z$ stays bounded between its initial value and zero, without
being allowed to change sign, until suddenly a flip takes place. The transition between
these two regimes can be seen in the evolution of $\Omega_e$ (bottom panel). At early times, $\Omega_e$
remains bounded between $-100^\circ$ and  $+100^\circ$. However, by the time $j_z$ changes sign,
$\Omega_e$ is circulating, sweeping all possible angles. An analogous plot of $\Omega_e$ for the example
in Figure~\ref{fig:case91} shows that $\Omega_e$ never circulates during the entire extent of the integration.

%%%%%%%%%%%%%%%%%%%%%%%%%%%%%%%%%%%%%%%%%
%%%%%%%%%%%%%%%%%%%%%%%%%%%%%%%%%%%%%%%%%
%%%%%%%%%%%%%%%%%%%%%%%%%%%%%%%%%%%%%%%%%
%%%%%%%%%%%%%%%%%%%%%%%%%%%%%%%%%%%%%%%%%

%%%%%%%%%%%%%%%%%%%%%%%%%%%%%%%%%%%%%%%%%
%%%%%%%%%%%%%%%%%%%%%%%%%%%%%%%%%%%%%%%%%
%%%%%%%%%%%%%%%%%%%%%%%%%%%%%%%%%%%%%%%%%
%%%%%%%%%%%%%%%%%%%%%%%%%%%%%%%%%%%%%%%%%

%%%%%%%%%%%%%%%%%%%%%%%%%%%%%
\subsection{Maximum eccentricity: analytical results}\label{sec:max_ecc}
%

%%%%%%%%%%%%%%%%%%%%%%%%%%%%%%%%%%%%%%%%%
%%%%%%%%%%%%%%%%%%%%%%%%%%%%%%%%%%%%%%%%%
\begin{figure*}
\begin{centering}
\includegraphics[width=5.7cm]{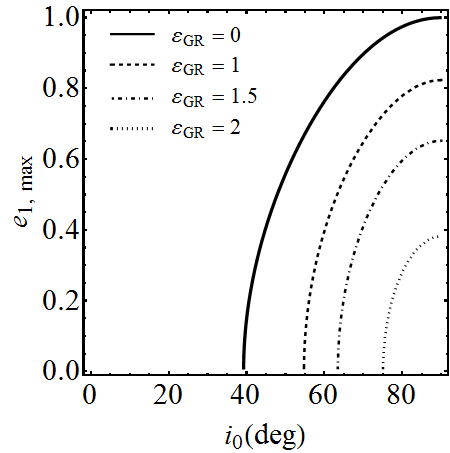}
\includegraphics[width=5.7cm]{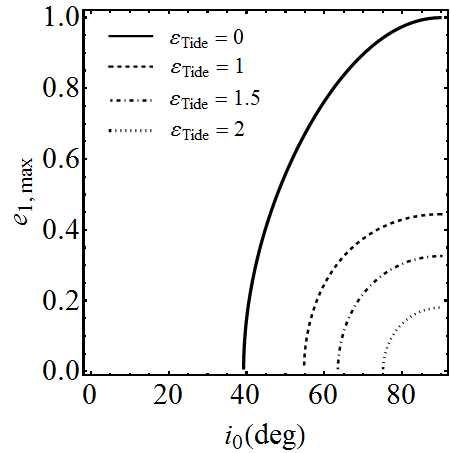}
\includegraphics[width=5.7cm]{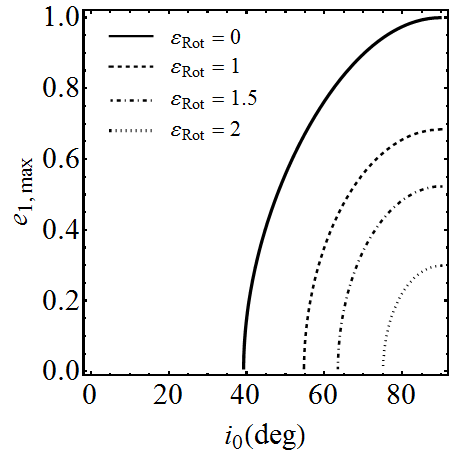}
\caption{The maximum eccentricity of the inner binary (when $m_1\ll m_0$) as a function the initial
inclination $i_0$ for different values of $\varepsilon_\gr$, $\varepsilon_\tide$, $\varepsilon_\rot$.
In these illustrative examples, SRFs compete with Lidov-Kozai oscillations from the start (when $e_{1,0}\sim0$);
however, for more realistic parameters (see Section~\ref{sec:paramspace} below), SRFs only dominate
over the tidal potential $\Phi$ when $e_1\sim1$ (see values of $\varepsilon_\mathrm{extra}$ in Table~\ref{tab:paramspace}).
}
\label{fig:emaxi}
\end{centering}
\end{figure*}
%%%%%%%%%%%%%%%%%%%%%%%%%%%%%%%%%%%%%%%%%
%%%%%%%%%%%%%%%%%%%%%%%%%%%%%%%%%%%%%%%%%

We see in Section 3.2 that SRFs limit the maximum eccentricity that can be achieved during the Lidov\---Kozai
cycles \citep[e.g.,][]{Holman,FT}. In the test-mass approximation ($m_1\ll m_0$) and
neglecting the octupole effect, this maximum eccentricity, $e_{1,\m}$, can be derived analytically.
We shall see in Section 4 that the limiting eccentricity, $e_\li$,
achieved for the initial inclination $i_0=90^\circ$, is also applicable when the octupole effect is included
and binaries of comparable masses ($m_1\sim m_0$) are considered.

In the test-mass approximation ($m_1\ll m_0$, which implies $d\jvec_2/dt=d\evec_2/dt=0$) and at the quadrupole level
($\varepsilon_\oct=0$), Equation~(\ref{eq:j1vec}) implies
\begin{equation}\label{eq:jConst}
\jvec_1\cdot\nvec_2=\sqrt{1-e_1^2}\cos\!i_1=\mathrm{constant}~~.
\end{equation}
In addition, the total potential is conserved
\begin{equation}\label{eq:total_potential}
\langle\Phi_{\mathrm{tot}}\rangle\approx\langle\Phi_{\K}\rangle+\langle\Phi_{\gr}\rangle
+\langle\Phi_{\tide}\rangle+\langle\Phi_{\rot}\rangle=\mathrm{constant}~~.
\end{equation}
In terms of the orbital elements of the inner binary,
the quadrupole potential (Equation~\ref{eq:APK}) can be written as
\begin{equation}\label{eq:APK2}
\langle\Phi_{\K}\rangle=-\frac{\mu_1\Phi_0}{8}\Big[2+3\e1-(3+12\e1-15\e1\cos^2\omega_1)\sin^2\I\Big],
\end{equation}

If the system is initialized with $e_1=0$, the maximum eccentricity $e_{1,\m}$ is achieved at
$\omega_1=\pi/2$ or $3\pi/2$ during the Lidov\---Kozai cycles.
Using Equations (\ref{eq:jConst}) (\ref{eq:APK2}), we find that $e_{1,\m}$ is given by
\begin{equation}\label{eq:FT}
\begin{split}
&\varepsilon_{\gr}\bigg(\frac{1}{j_{1,\mi}}-1\bigg)
+\frac{\varepsilon_{\tide}}{15}\bigg(\frac{1+3e^2_{1,\m}+\frac{3}{8}e^4_{1,\m}}{j_{1,\mi}^9}-1\bigg)\\
&+\frac{\varepsilon_{\rot}}{3}\bigg(\frac{1}{j_{1,\mi}^3}-1\bigg)
=\frac{9}{8}\frac{e^2_{1,\m}}{j_{1,\mi}^2}\bigg(j_{1,\mi}^2-\frac{5}{3}\cos^2\!i_0\bigg)~,
\end{split}
\end{equation}
where $j_{1,\mi}=\sqrt{1-e^2_{1,\m}}$. In the absence of the SRFs
($\varepsilon_\gr=\varepsilon_\tide=\varepsilon_\rot=0$),
the above equation yields the well-known maximum eccentricity $e_{\mathrm{m}0}$
for ``pure" Lidov\---Kozai oscillation \citep[e.g.,][]{Kozai,Lidov}
\begin{equation}\label{eq:ME}
e_{\mathrm{m}0}= \sqrt{1-\frac{5}{3}\cos^2i_0}~,
\end{equation}
If we neglect the tidal or rotation terms ($\varepsilon_\tide=\varepsilon_\rot=0$)
and assume $\varepsilon_\gr\ll1$, Equation~(\ref{eq:FT}) results to \citep[e.g.,][]{MH}
\begin{equation}
j_{1,\mi}=\frac{1}{9}\Big[4\varepsilon_\gr+\sqrt{16\varepsilon_\gr^2+135\cos^2\!i_0}\Big]
\end{equation}

Figure \ref{fig:emaxi} depicts several example of $e_{1,\m}$ for different values of
$\varepsilon_\gr$, $\varepsilon_\tide$, $\varepsilon_\rot$.

The physical meaning of Equation~(\ref{eq:FT}) can be made clear if we use the expression of
$\Dot \omega_\extra$ Equations \ref{eq:omegagr}\---\ref{eq:omegarot}) to re-express it as
(assuming $j_{1,\mi}\ll1$)
\begin{equation}
\begin{split}
&\Bigg[\frac{\Dot\omega_\gr}{\Dot\omega_K}+\frac{1}{15}\frac{\Dot\omega_\tide}{\Dot\omega_K}f(e_1)
+\frac{1}{3}\frac{\Dot\omega_\rot}{\Dot\omega_K}\Bigg]_{e_1=e_{1,\m}}\\
&\approx\frac{9}{8}e_{1,\m}^2\frac{j_{1,\mi}^2-5\cos^2\!i_0/3}{j_{1,\mi}^2}~,
\end{split}
\end{equation}
where
\begin{equation}
f(e_1)\equiv\frac{1+3\e1+\frac{3}{8}e_1^4}{1+\frac{3}{2}\e1+\frac{1}{8}e_1^4}~,
\end{equation}
{and where we have defined for convenience a ``characteristic" Lidov\---Kozai rate}
\begin{equation}
\Dot\omega_K\equiv\frac{1}{t_K\sqrt{1-\e1}}~,
\end{equation}
{which should not be confused with the precession rate derived from applying the operation
in Equation~\ref{eq:extraomega} to $\langle\Phi_{\K}\rangle$.}

For $i_0=90^\circ$, the eccentricity attains the limiting value, $e_\li\equiv e_{1,\m}$, given by
\begin{equation}\label{eq:FTT}
\Bigg[\frac{\Dot\omega_\gr}{\Dot\omega_K}+\frac{1}{15}\frac{\Dot\omega_\tide}{\Dot\omega_K}f(e_1)
+\frac{1}{3}\frac{\Dot\omega_\rot}{\Dot\omega_K}\Bigg]_{e_1=e_\li}=\frac{9}{8}e_\li^2~.
\end{equation}

For $1-e_\li\ll1$, we have $f(e_\li)\simeq5/3$, Equation~(\ref{eq:FTT}) becomes
\begin{equation}\label{eq:L}
\Bigg[\frac{\Dot\omega_\gr}{\Dot\omega_K}+\frac{1}{9}\frac{\Dot\omega_\tide}{\Dot\omega_K}
+\frac{1}{3}\frac{\Dot\omega_\rot}{\Dot\omega_K}\Bigg]_{e_1=e_\li}\simeq\frac{9}{8}~.
\end{equation}
Thus, the limiting eccentricity is achieved when the periapse precession rate due to
SRFs becomes comparable to the Lidov\---Kozai rate $\Dot \omega_K$.

%%%%%%%%%%%%%%%%%%%%%%%%%%%%%%%%%%%%%%%%%%%%%%%%%%%%%%%%%%%
\section{Parameter Survey: Test-Mass Cases}\label{sec:paramspace}
In the test-mass limit ($m_1\ll m_0$),
the evolution of the inner binary depends on the dimensionless ratios $\varepsilon_\oct$,
$\varepsilon_\gr$, $\varepsilon_\tide$, $\varepsilon_\rot$ as well as the initial inclination
angle $i_0$ (we assume $e_0\simeq0$).
In this section, we consider the evolution of Jupiter-mass planet ($m_1=m_J, R_1=R_J$)
moving around a Solar-mass star ($m_0=m_\odot$).
We carry out calculations for different values of $a_1$, $a_2$, $m_2$ and $e_2$.
The different orbital configurations and their corresponding values of 
$\varepsilon_\oct$ and $\varepsilon_\mathrm{extra}$ are listed in Table~\ref{tab:paramspace}.
These conditions of parameters are subject to the stability criterion of \citet{MA}
\begin{equation}\label{eq:MA}
\frac{a_2}{a_1}>2.8\bigg(1+\frac{m_2}{m_0}\bigg)^{2/5}
\frac{(1+e_2)^{2/5}}{(1-e_2)^{6/5}}\bigg(1-\frac{0.3\I}{180^\circ}\bigg)~~.
\end{equation}
%

%%%%%%%%%%%%%%%%%%%%%%%%%%%%%%%%%%%%%%%%%%%%%%%%%%%%%%%%%%%%%
%%%%%%%%%%%%%%%%%%%%%%%%%%%%%%%%%%%%%%%%%%%%%%%%%%%%%%%%%%%%%
%%%%%%%%%%%%%%%%%%%%%%%%%%%%%%%%%%%%%%%%%%%%%%%%%%%%%%%%%%%%%
\begin{table*}
 \centering
 \begin{minipage}{120mm}
  \caption{Initial conditions on different cases: $m_0=1M_\odot$, $m_1=1M_J$, $e_0=0.001$, $k_{2,1}=0.37$, $k_{q,1}=0.17$, $R_1=1R_J$, {$\omega_1=\omega_2=\Omega_1=0^\circ$ and $\Omega_2=180^\circ$.}
  The parameter $\varepsilon_\extra$ is calculated by definition in Equations~(\ref{eq:R11}), (\ref{eq:R12}) and (\ref{eq:R13}).
\label{tab:paramspace}
  }
  \begin{tabular}{@{}llrrrrlrlr@{}}
  \hline
   Parameter & $\varepsilon_{\oct}$ & $\varepsilon_{\gr}$ ~~~~~& $\varepsilon_{\tide}$~~~~~
     & $a_1 (\au)$ & $a_2(\au)$ & $e_2$~ & $m_2(M_\odot)$~~~~~  \\
 \hline
 Case 1 & 0.001 & $4.47\times10^{-1}$ & $2.61\times10^{-6}$~   & 1~~~~~ & 200~~~ & 0.2 & 0.5~~~~~~~~~  \\
 \\
 Case 2 & 0.002 & $2.79\times10^{-2}$ & $1.63\times10^{-7}$~   & 1~~~~~ & 100~~~ & 0.2 & 1~~~~~~~~~~  \\
 \\
 Case 3a & 0.006 & $1.72\times10^{-4}$ & $7.78\times10^{-13}$ & 6~~~~~ & 200~~~ & 0.2 & 1~~~~~~~~~~  \\
 Case 3b & 0.006 & $1.03\times10^{-3}$ & $6.05\times10^{-9}$~  & 1~~~~~ & 33.33~~ & 0.2 & 1~~~~~~~~~~  \\
 Case 3c & 0.006 & $1.03\times10^{-2}$ & $6.05\times10^{-8}$~  & 1~~~~~ & 33.33~~ & 0.2 & 0.1~~~~~~~~~  \\
 \\
 Case 4 & 0.011 & $5.13\times10^{-2}$ & $3.00\times10^{-7}$~  & 1~~~~~ & 200~~~ & 0.8 & 1~~~~~~~~~~  \\
 \\
 Case 5 & 0.013 & $2.15\times10^{-5}$ & $9.72\times10^{-14}$  & 6~~~~~ & 100~~~ & 0.2 & 1~~~~~~~~~~  \\
 \\
 Case 6 & 0.022  & $6.41\times10^{-3}$ & $3.75\times10^{-8}$~  & 1~~~~~ & 100~~~ & 0.8 & 1~~~~~~~~~~  \\
 \\
 Case 7 & 0.033  & $3.08\times10^{-5}$ & $2.89\times10^{-13}$   & 5~~~~~ & 100~~~ & 0.5 & 1~~~~~~~~~~ \\
 \\
 Case 8 & 0.044  & $4.01\times10^{-4}$ & $1.46\times10^{-10}$    & 2~~~~~ & 100~~~ & 0.8 & 1~~~~~~~~~~ \\
 \\
 Case 9 & 0.056  & $2.93\times10^{-4}$ & $1.32\times10^{-12}$   & 6~~~~~ & 100~~~ & 0.6 & 0.04~~~~~~~~ \\
 \\
 Case 10a & 0.067 & $3.96\times10^{-5}$ & $1.79\times10^{-13}$ & 6~~~~~ & 200~~~ & 0.8 & 1~~~~~~~~~~  \\
 Case 10b & 0.067 & $2.37\times10^{-4}$ & $1.39\times10^{-9}$~ & 1~~~~~ & 33.33~~ & 0.8 & 1~~~~~~~~~~  \\
 Case 10c & 0.067 & $2.37\times10^{-3}$ & $1.39\times10^{-8}$~  & 1~~~~~ & 33.33~~ & 0.8 & 0.1~~~~~~~~~  \\
 \\
 Case 11 & 0.133  & $4.95\times10^{-6}$ & $2.23\times10^{-14}$  & 6~~~~~ & 100~~~ & 0.8 & 1~~~~~~~~~~  \\
\hline
\end{tabular}
\end{minipage}
\end{table*}
%%%%%%%%%%%%%%%%%%%%%%%%%%%%%%%%%%%%%%%%%%%%%%%%%%%%%%%%%%%%%
%%%%%%%%%%%%%%%%%%%%%%%%%%%%%%%%%%%%%%%%%%%%%%%%%%%%%%%%%%%%%
%%%%%%%%%%%%%%%%%%%%%%%%%%%%%%%%%%%%%%%%%%%%%%%%%%%%%%%%%%%%%

%%%%%%%%%%%%%%%%%%%%%%%%%%%%%%%%%%%%%%%%%%%%%%%%%%%%%%%%%%%%%
%%%%%%%%%%%%%%%%%%%%%%%%%%%%%%%%%%%%%%%%%%%%%%%%%%%%%%%%%%%%%
%%%%%%%%%%%%%%%%%%%%%%%%%%%%%%%%%%%%%%%%%%%%%%%%%%%%%%%%%%%%%
\begin{table*}
 \centering
 \begin{minipage}{170mm}
  \caption{Results for the  different cases listed in Table~1.
The limiting eccentricity  $e_\li$ is obtained from Equation~(\ref{eq:FTT}) in the case of $i_0=90^\circ$ and pseudo synchronized.
 $e_{\m,{\mathrm{Num}}}$ means the maximum eccentricity found from the numerical integration's results.
We also define three angles:
$i_0|_{e\li}^{\mathrm{SRF}}$ is the initial inclination where the simulation result $e_\m$ first reaches the analytic value $e_\li$,
$i_0|_{\mathrm{flip}}$ is the smallest angle at which the first flipping orbit occurs without tidal friction
and $i_0|_{\mathrm{flip}}^{\mathrm{SRF}}$ is the one with SRFs.
          }
  \begin{tabular}{@{}llrrrrlrlr@{}}
\hline
   Parameter &  ~~~$1-e_\li$ & $1-e_{\m,{\mathrm{Num}}}$  & $\left.\frac{\Dot \omega_\gr}{\Dot \omega_\mathrm{K}}\right|_{e_\li} $
      & $\left.\frac{\Dot \omega_\tide}{\Dot \omega_\mathrm{K}}\right|_{e_\li}$
      & $\left.\frac{\Dot \omega_\rot}{\Dot \omega_\mathrm{K}}\right|_{e_\li}$
      & $\left.i_0\frac{}{}\right|_{\mathrm{flip}}$(deg) & $\left.i_0\frac{}{}\right|_{\mathrm{flip}}^{\mathrm{SRF}}$(deg)  & $\left.i_0\frac{}{}\right|_{e\li}^{\mathrm{SRF}}$(deg)\\
 \hline
 Case 1 & $5.01\times10^{-2}$ & $4.92\times10^{-2}$ ~ &  $1.43$~~~~~~  & $2.24\times10^{-1}$ & $2.76\times10^{-2}$  & ~~~~~~87.7  & $\sim90$~~~~~~~& ~~~~88.5    \\
 \\
 Case 2 & $1.28\times10^{-2}$ & $1.11\times10^{-2}$ ~ &  $1.75\times10^{-1}$  & 6.23~~~~~~ &  $7.85\times10^{-1}$  & ~~~~~~86.2 & $\sim90$~~~~~~ &  ~~~~85.4   \\
 \\
 Case 3a & $8.12\times10^{-4}$ & $7.00\times10^{-4}$ ~  & $4.28\times10^{-3}$  & 7.29~~~~~~ &  $9.27\times10^{-1}$& ~~~~~~82.2 & $\sim90$~~~~~~ &  ~~~~87.1    \\
 Case 3b & $5.97\times10^{-3}$ & $5.16\times10^{-3}$ ~ & $9.48\times10^{-3}$  & 7.21~~~~~~ &  $9.13\times10^{-1}$& ~~~~~~82.2 & $\sim90$~~~~~~  &  ~~~~86.8    \\
 Case 3c & $1.01\times10^{-2}$ & $8.71\times10^{-3}$ ~  & $7.30\times10^{-2}$  & 6.81~~~~~~ &  $8.60\times10^{-1}$& ~~~~~~82.2 & $\sim90$~~~~~~ &   ~~~~85.4    \\
 \\
 Case 4 & $1.51\times10^{-2}$ & $1.31\times10^{-4}$ ~ & $2.91\times10^{-1}$  & 5.56~~~~~~ &  $7.00\times10^{-1}$& ~~~~~~80.2 & $\sim90$~~~~~~ &   ~~~~83.9    \\
 \\
 Case 5 & $5.11\times10^{-4}$ & $4.41\times10^{-4}$ ~ & $6.74\times10^{-4}$  & 7.32~~~~~~~&  $9.31\times10^{-1}$ & ~~~~~~79.0 & 89.4~~~~~~ &  ~~~~78.8    \\
 \\
 Case 6 &$9.02\times10^{-3}$  & $7.78\times10^{-3}$ ~ &  $4.78\times10^{-2}$  & 6.96~~~~~~ &  $8.80\times10^{-1}$  & ~~~~~~69.9 & 89.1~~~~~~ &   ~~~~71.9    \\
 \\
 Case 7 & $6.51\times10^{-4}$  & $5.51\times10^{-4}$ ~  &  $8.55\times10^{-4}$  & 7.32~~~~~~ &  $9.30\times10^{-1}$& ~~~~~~58.7 & 60.4~~~~~~ &   ~~~~59.3    \\
 \\
 Case 8 & $2.60\times10^{-3}$  & $2.23\times10^{-3}$ ~  &  $5.55\times10^{-3}$  & 7.27~~~~~~ &  $9.23\times10^{-1}$& ~~~~~~51.3 & 73.3~~~~~~ &   ~~~~51.9    \\
 \\
 Case 9 & $9.14\times10^{-4}$  & $7.71\times10^{-4}$ ~  &  $6.86\times10^{-3}$  & 7.28~~~~~~ &  $9.25\times10^{-1}$& ~~~~~~47.8 & 56.0~~~~~~ &   ~~~~49.1    \\
 \\
 Case 10a &$5.85\times10^{-4}$ &$4.91\times10^{-4}$ ~ & $1.16\times10^{-3}$ & 7.31~~~~~~ &  $9.30\times10^{-1}$& ~~~~~~47.3 & 53.9~~~~~~  &  ~~~~47.8    \\
 Case 10b & $4.29\times10^{-3}$ & $3.67\times10^{-3}$ ~ & $2.56\times10^{-3}$  & 7.27~~~~~~ &  $9.22\times10^{-1}$& ~~~~~~47.3 & 73.1~~~~~~ &   ~~~~51.9    \\
 Case 10c & $7.19\times10^{-3}$ & $6.15\times10^{-3}$ ~ & $1.98\times10^{-2}$  & 7.14~~~~~~ &  $9.03\times10^{-1}$& ~~~~~~47.3 & 81.1~~~~~~ &  ~~~~52.1     \\
 \\
 Case 11 & $3.68\times10^{-4}$  & $3.09\times10^{-4}$ ~  & $1.82\times10^{-4}$  & 7.32~~~~~~ &  $9.31\times10^{-1}$& ~~~~~~47.3 & 48.4~~~~~~ &   ~~~~49.6    \\
\hline
\end{tabular}
\end{minipage}
\end{table*}
%%%%%%%%%%%%%%%%%%%%%%%%%%%%%%%%%%%%%%%%%%%%%%%%%%%%%%%%%%%%%
%%%%%%%%%%%%%%%%%%%%%%%%%%%%%%%%%%%%%%%%%%%%%%%%%%%%%%%%%%%%%
%%%%%%%%%%%%%%%%%%%%%%%%%%%%%%%%%%%%%%%%%%%%%%%%%%%%%%%%%%%%%

%%%%%%%%%%%%%%%%%%%%%%%%%%%%%%%%%%%%%%%%%
%%%%%%%%%%%%%%%%%%%%%%%%%%%%%%%%%%%%%%%%%
\begin{figure*}
\begin{centering}
\includegraphics[width=12cm]{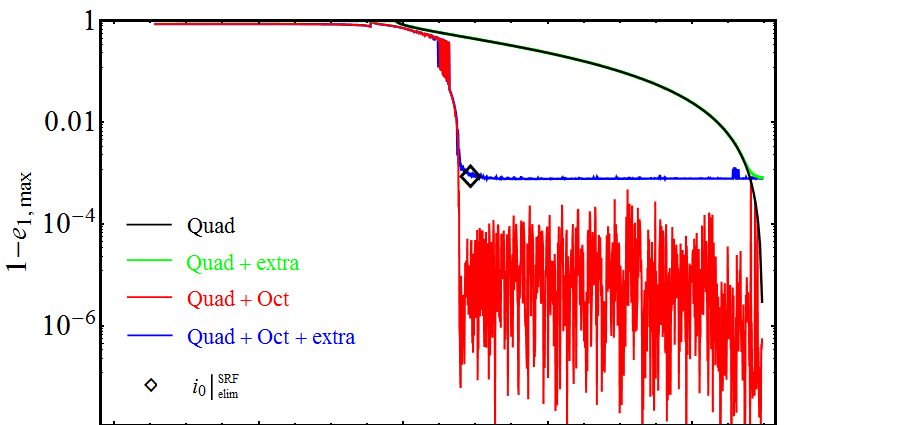}
\includegraphics[width=12cm]{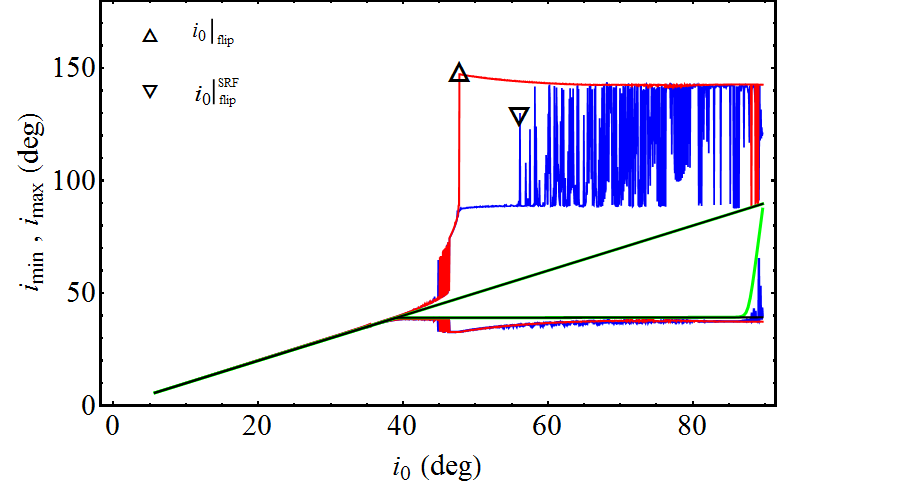}
\caption{Case 9 in Table 1 with
$\varepsilon_{\oct}=0.056$. Same values of $m_0$, $m_1$, $m_2$, $a_1$, $a_2$
and $e_2$ from Figures~\ref{fig:case91} and~\ref{fig:case92},
extending initial mutual inclination to the full range $(0^\circ,90^\circ)$
{(in all cases,  $\omega_1=0^\circ$, $\omega_2=0^\circ$, $\Omega_1=0^\circ$ and $\Omega_2=180^\circ$ at $t=0$).} We integrate Equations (\ref{eq:fulle1})-(\ref{eq:fullomega2}) for quadrupole ($\varepsilon_\oct=0$) 
and octupole ($\varepsilon_\oct\neq0$) approximations of the potential as well as without ($\varepsilon_\mathrm{extra}=0$) 
and with ($\varepsilon_\mathrm{extra}\neq0$) SRFs. The total integration time is $5\times10^{7}$ years ($\sim 360.5~t_K$).
The upper panel shows the maximum eccentricity $e_{1,\m}$ achieved over the entire integration time for the four different approximation
used. Similarly, the lower panel shows the extrema in inclination $i_{0,\m/\mi}$ attained during the evolution. 
Black curves correspond to ``pure" quadrupole-level Lidov\---Kozai oscillations (i.e., $\varepsilon_\oct=0$ and $\varepsilon_\mathrm{extra}=0$);
green curves correspond to quadrupole-level Lidov\---Kozai oscillations {\it with} SRFs 
($\varepsilon_\oct=0$ and $\varepsilon_\mathrm{extra}\neq0$); red curves correspond to
``pure" octupole-level Lidov\---Kozai oscillations ($\varepsilon_\oct\neq0$ and $\varepsilon_\mathrm{extra}=0$); and
blue curves correspond to octupole-level Lidov\---Kozai oscillations with SRFs ($\varepsilon_\oct\neq0$ and $\varepsilon_\mathrm{extra}\neq0$).
Blue curves show how strict SRFs are in establishing a global maximum eccentricity, capping the octupole-level evolution
at the limiting value $e_\li$ given by Equation~\ref{eq:FTT}.  Inclination is strongly affected by the octupole terms
even in the presence of SRFs, however flips are suppressed $i_0$ below $60^\circ$, limiting those orbits to a maximum
inclination of $90^\circ$, but not beyond that value. Triangular and square markers denote different transitions in
the behaviour of $e_1$ and $i_1$ (see text and Table~2).}
\label{fig:case94}
\end{centering}
\end{figure*}
%%%%%%%%%%%%%%%%%%%%%%%%%%%%%%%%%%%%%%%%%
%%%%%%%%%%%%%%%%%%%%%%%%%%%%%%%%%%%%%%%%%

%%%%%%%%%%%%%%%%%%%%%%%%%%%%%%%%%%%%%%%%%
%%%%%%%%%%%%%%%%%%%%%%%%%%%%%%%%%%%%%%%%%
\begin{figure*}
\begin{centering}
\includegraphics[width=12cm]{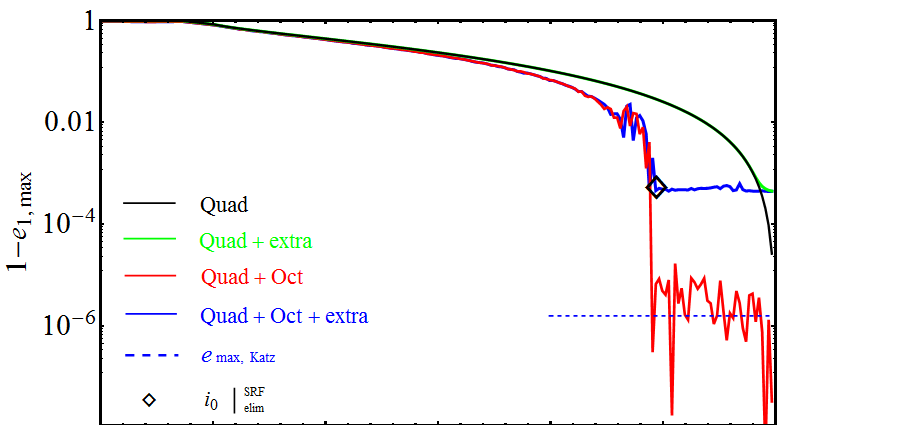}
\includegraphics[width=12cm]{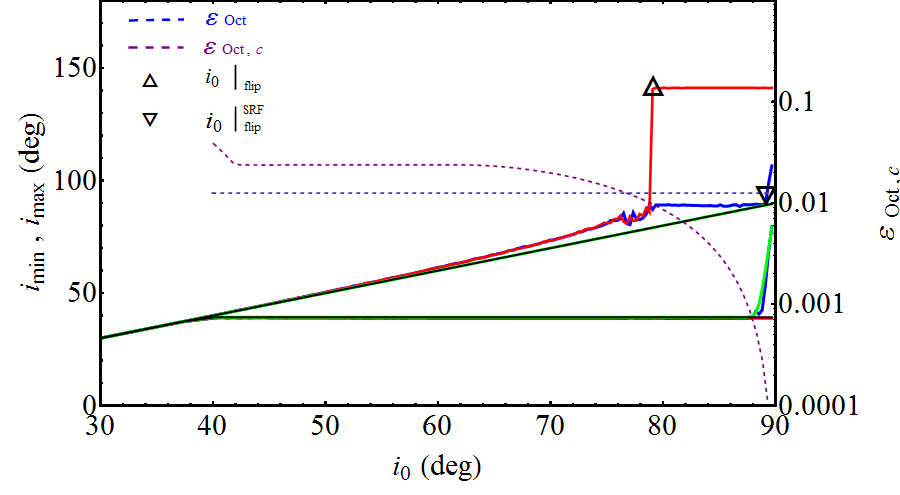}
\caption{Same as Figure~\ref{fig:case94}, but now corresponding to Case 5 in Table 1.
($\varepsilon_{\oct}=0.013$). We integrate the Equations for $5\times10^{6}$ years ($\sim 490.6~t_K$).
System parameters are $m_0=1M_\odot$, $m_1=1M_J$,  $m_2=1M_\odot$, $a_1=6\au$ and $a_2=100\au$.
We start each run with $e_1=0.001$, $e_2=0.2$,
{$\omega_1=0^\circ$, $\omega_2=0^\circ$, $\Omega_1=0^\circ$ and $\Omega_2=180^\circ$.}
 Qualitative behaviour of this set of systems is similar to
that of Figure~\ref{fig:case94}, except that now the ``window of influence" of octupole effects
is much narrower, limiting the importance of octupolar corrections and enabling
SRFs to severely limit their resulting extreme orbits. In particular, although octupole terms can still alter
inclinations beyond the quadrupole-level predictions, in this example orbital flips have been entirely
suppressed. Dashed lines correspond to the estimates provided by \citet{Katz PRL} for
the maximum eccentricity $e_\mathrm{max,Katz}$ within the ``window of influence" (upper panel) and
for the critical value $\varepsilon_{\oct,c}$ above which orbital flips are allowed (see Section 4.3).
}
\label{fig:case5}
\end{centering}
\end{figure*}
%%%%%%%%%%%%%%%%%%%%%%%%%%%%%%%%%%%%%%%%%
%%%%%%%%%%%%%%%%%%%%%%%%%%%%%%%%%%%%%%%%%

For each combination of $\varepsilon_\oct$ and $\varepsilon_\mathrm{extra}$ (Table~\ref{tab:paramspace}),
we integrate a total of 300 triple systems over a total integration time ranging from $\sim360t_K$ to $\sim500t_K$. 
We setup each system by varying the initial inclination of the inner binary $i_0$ ($\simeq i_\mathrm{tot,0}$ when
$m_1\ll m_0$) between $0^\circ$ and $90^\circ$.

%%%%%%%%%%%%%%%%%%%%%%%%%%%%%%%%%%%%
\subsection{A fiducial example with $\varepsilon_\oct=0.056$}
We first consider a specific example with $\varepsilon_\oct=0.056$ (Case 9 in Table 1).
For each initial inclination angle $i_0$ (in the range between $0^\circ$ and $90^\circ$),
we integrate Equations (\ref{eq:fulle1})-(\ref{eq:fullomega2}) for $5\times10^7$ yrs,
corresponding to 360.5 $t_K\sim 10t_K/\varepsilon_\oct$ for this specific set of parameters.
For each of these subsystems, we record the maximum eccentricity $e_{1,\m}$ and the maximum and minimum of $\I$ attained
during the evolution. The results are shown in Figure \ref{fig:case94}.

%%%%%%%%%%%%%%%%%%%
\subsubsection{Eccentricity maxima}
The upper panel in Figure \ref{fig:case94} shows the maximum eccentricity of inner orbit
as a function of $i_0$.
At the quadrupole level,
the pure Lidov\---Kozai cycles can give extremely large eccentricity ($1-e_{1,\m}\leq10^{-5}$)
only for $i_0\simeq90^\circ$.
When the short-range effects are included,
the value of $e_{1,\m}$ at that point is limited to $1-e_{1,\m}\leq10^{-3}$,
in accordance with the analytic expression (Equation \ref{eq:FT}).
For the parameters considered in this case (Case 9),
the tidal effect plays the dominate role in limiting the maximum eccentricity
(Table 2, when we see that $\Dot \omega_\tide/(9\Dot \omega_K)\gg\Dot \omega_\rot/(3\Dot \omega_K)\gg\Dot \omega_\gr/\Dot \omega_K$
at $e_1=e_\li$; see also Equation \ref{eq:L}).

When the octupole term is included, there is a sharp jump of $e_{1,max}$
at $i_0\approx 50^\circ$. Without the SRFs, $(1-e_{\rm max})$ becomes very small
and varies erratically as $i_0$ increases beyond $50^\circ$.
This erratic variation is the result of the overlap between the
quadrupole and octupole contributions.  When the SRFs are included, we find
that instead of the rapid variation of $(1-e_{\rm max})$,
the maximum eccentricity becomes approximately a constant, equal to $e_{\rm lim}$.
We define $i_0|_{e\li}^{\mathrm{SRF}}$ as the the value of $i_0$ when $e_{\rm
 max}$ first reaches $e_{\rm lim}$. For the case considered in
Fig.~7, $i_0|_{e\li}^{\mathrm{SRF}}$ is close to $50^\circ$.
It is important to note that although $e_{\rm lim}$ is derived
in the quadrupole approximation (see Section 3.3), it serves as the maximum
eccentricity attainable even when the octupole term is included.

%%%%%%%%%%%%%%%%%%%
\subsubsection{Inclination extrema}
The lower panel in Figure \ref{fig:case94} shows the maximum and minimum of
the orbital inclination attained during the evolution of the inner
binary as functions of $i_0$.
At the pure quadrupole level (without SRFs),
the inclination does not change until $i_0$ reaches $\arccos\sqrt{3/5}\simeq 40^\circ$,
beyond which $i_{\rm max}=i_0$ and $i_{\rm min}=40^\circ$. Including SRFs, $i_{\rm min}$
is modified for $i_0$ close to $90^\circ$ as the maximum eccenticity is limited by the SRFs.
Note that $\sqrt{1-e_{\rm max}^2}\cos i_{\rm min}=\cos i_0$.

At the octupole level, the angular momentum of the inner binary
experiences a flip ($i_{\rm max}>90^\circ$) at a critical angle
$i_0|_{\mathrm{flip}}$.  This angle is always greater than $40^\circ$
(the onset of quadrupole Lidov\---Kozai oscillations). When the SRFs are
included, this critical angle is pushed to a higher value,
$i_0|_{\mathrm{flip}}^{\mathrm{SRF}}$.

At inclinations slightly higher than the angle critical $i_0|_{e\li}^{\mathrm{SRF}}$
(demarcated by inverted triangle in Figure~\ref{fig:case94}, the orbital flips seen in absence of SRFs
(red curves) have now been inhibited. Note, however, that at even higher inclinations
( $i_0> i_0|_{\mathrm{flip}}^{\mathrm{SRF}}$), orbital flips are once again
allowed. This can occur even though the maximum eccentricity  $e_{1,\m}$ is
strongly affected by the SRFs (see Figure~\ref{fig:case93} and Section~\ref{sec:examples}).
Note also that the individual examples shown in Figures~\ref{fig:case92} and~\ref{fig:case93}
(with $i_0$ of $65^\circ$ and $65.3783^{\circ}$ respectively) lie
within the erratically varying region that starts at $i_0|_{\mathrm{flip}}^{\mathrm{SRF}}\sim56^\circ$.
It is possible that the inclination angle in the example with $i_0=65^\circ$ (Figure~\ref{fig:case92}) will eventually
flip as well. From this figure we can conclude that SRFs have broken the symmetry that existed between
the upper and lower panels of Figure~\ref{fig:case94} (red curves), which showed that
both the extreme maximum eccentricity and the orbital flip where achieved above the same
critical inclination.

The erratic variation of $i_{\rm max}$ for $i_0>
i_0|_{\mathrm{flip}}^{\mathrm{SRF}}$ results from the combined effects
of quadrupole, octupole and SRFs, and may be associated with the chaotic
behaviour of Lidov\---Kozai oscillations studied by \citet{Li 2014}.
These authors find that configurations with higher inclinations and larger
$\varepsilon_\oct$ are more chaotic (with Lyapunov times of $\sim6t_K$ for the chaotic
regions of parameter space).  It is possible that the
complexity of a system with conservative SRFs (three additional frequencies are present
and no energy dissipation) only shift the inclination threshold for chaotic behaviour to larger angles,
but have not fundamentally suppressed the chaotic nature of Lidov\---Kozai oscillations
with octupole-level terms. It is also possible that the characteristic timescale for a flip has been
entirely altered by the SRFs, and that all systems with $i_0>
i_0|_{\mathrm{flip}}^{\mathrm{SRF}}$ will eventually flip (on timescales much longer
than $t_K/\varepsilon_\oct$).

%%%%%%%%%%%%%%%%%%%%%%%%%%%%%%%
\subsection{Dependence of $e_{1,\mathrm{max}}$  and $i_0|_{\mathrm{flip}}$ with $\varepsilon_\oct$}
To explore how the extrema in $e_1$ and $i_1$ change with
$i_0$ for different values of $\varepsilon_\oct$,
we carry out another set of numerical
integrations, this time with  $\varepsilon_\oct=0.013$ (listed as `Case 5' in Table~1 below).
The results of this set of integrations are shown in Figure~\ref{fig:case5}. Without SRFs (red curves),
the  $e_{1,\mathrm{max}}$  and  $i_{\mathrm{min}/\mathrm{max}}$ curves
exhibit the same overall morphology observed in the case with $\varepsilon_\oct=0.056$
 (Figure~\ref{fig:case94}).
In this case, however, significant deviations from the quadrupole-only calculations
are confined to a narrower range in $i_0$. This is to be expected,
since this ``octupole active" region will gradually shrink as $\varepsilon_\oct$ is made smaller,
 until the quadrupole level solutions (black and green curves) are recovered.
  In the limit  $\varepsilon_\oct\rightarrow0$, the only angle
which allows for $e_{1,\mathrm{max}}=1$ is $i_0=90^\circ$ (Equation~\ref{eq:ME}).
Similarly, when $\varepsilon_\oct\rightarrow0$ only $i_0=90^\circ$ permits $j_z=0$.

Quantitatively, the width along the
$i_0$-axis of the octupole-active region or ``window of influence" for a given value of $\varepsilon_\oct$
can be understood using
the ``flip condition"  identified by \citet{Katz PRL}. From approximate conservation laws,
these authors find that, given $e_{1,0}\sim0$ and $j_{z,0}\sim\cos i_0$, the long
term oscillation of $j_z$ owing to octupole terms can only result in a change of sign
if and only if $i_0$ is greater than a critical value that depends on $\varepsilon_\oct$. Equivalently,
given $i_0$, there is a critical value $\varepsilon_{\oct,c}$ above which orbits will flip.
This  flip condition can be approximately expressed as
$\varepsilon_{\oct,c}=\tfrac{1}{2}F(\cos^2i_0/2)$ where $F(x)$ is a non-monotonic function
that is equal to zero at $x=0$ and $x\approx0.236$ and peaks at 0.0475 for $x\approx0.112$
\citep[see ][Eqs. 17]{Katz PRL}.
The critical value $\varepsilon_{\oct,c}$ is a monotonically decreasing function of $i_0$, meaning
that the closer $i_0$ is to $90^\circ$, the smaller $\epsilon_{\oct,c}$ becomes (i.e., the easier
it is to flip). We illustrate this by overlaying $\epsilon_{\oct,c}$ as a function of $i_0$
 in the bottom panel of  Figure~\ref{fig:case5}. When  $\epsilon_{\oct,c}$
 becomes smaller than $\varepsilon_{\oct}=0.013$ (at $i_0\sim77^\circ$),
 test particle trajectories are allowed to flip orientations. Above this critical angle,
 each orbital flip is accompanied by an extreme increase in eccentricity. Following
 \citet{Katz PRL}, we estimate that this limiting eccentricity within the octupole active region
 is such that $1-e_\mathrm{max,Katz}^2\approx(0.14~\varepsilon_\oct)^2$, i.e.,
 $1-e_\mathrm{max,Katz}\approx1.7\times10^{-6}$,
 which is in good agreement with the average value of $e_\mathrm{1,max}$ in the
 region where $i_0>79^\circ$.

When SRFs are included, (blue and green curves in Figure~\ref{fig:case5}), the modifications to the evolution
of eccentricity are consistent with what was observed in the $\varepsilon_\oct=0.056$  example.
However, the amplitude of the inclination oscillations
is more dramatically affected. On one hand, we find
a consistently truncated maximum eccentricity down to  a value of
$1-e_{1,\mathrm{max}}\approx4.41\times10^{-4}$,
in rough agreement with the value of $1-e_{\mathrm{lim}}\approx5.11\times10^{-4}$ predicted
by Equation~(\ref{eq:FTT}). On the other hand, the orbital flips above $i_0\sim79^\circ$ are
entirely suppressed, in contrast with the behaviour observed in Figure~\ref{fig:case94}, where
only a fraction of the systems have their orbital flips entirely suppressed, while at high
inclinations the orbits still manage to reverse their orientations despite the strict
limits on the maximum eccentricity.

%%%%%%%%%%%%%%%%%%%%%%%%%%%%%%%%%%%%%%%%%%%%%%
%%%%%%%%%%%%%%%%%%%%%%%%%%%%%%%%%%%%%%%%%%%%%%
\begin{figure}
\begin{centering}
\includegraphics[width=0.5\textwidth]{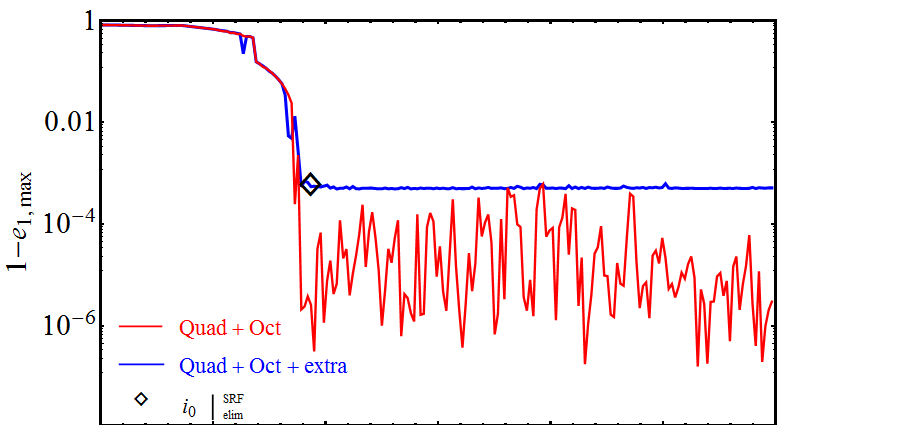}
\includegraphics[width=0.5\textwidth]{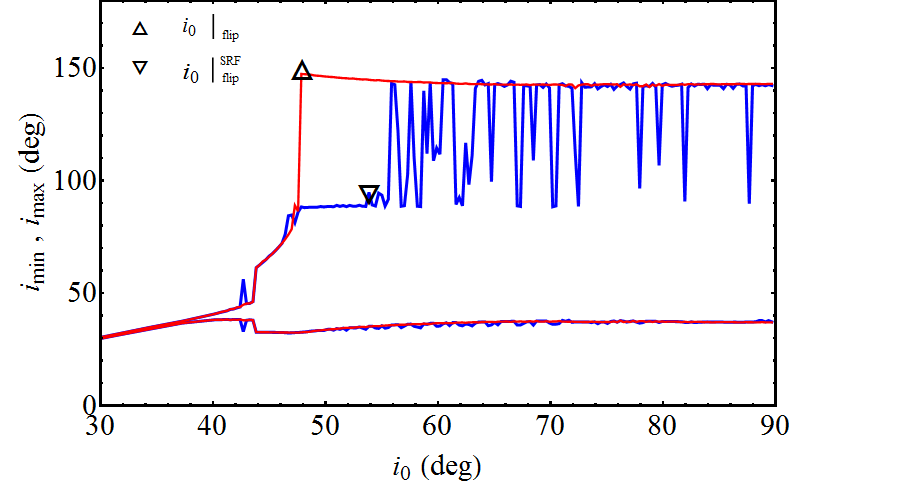}
\caption{Case 10a in Table 1 ($\varepsilon_{\oct}=0.067$,
$\varepsilon_\gr=3.96\times10^{-5}$, $\varepsilon_\tide=1.79\times10^{-13}$).
System parameters are $m_0=1M_\odot$, $m_1=1M_J$, $m_2=1M_\odot$
 $a_1=6\au$ and $a_2=200\au$. Orbits are started with $e_1=0.001$, $e_2=0.8$,
 {$\omega_1=0^\circ$, $\omega_2=0^\circ$, $\Omega_1=0^\circ$ and $\Omega_2=180^\circ$.}
The total integration time is $6\times10^{6}$ years ($\sim 320.5~t_K$).
}
\label{fig:case10a}
\end{centering}
\end{figure}
%%%%%%%%%%%%%%%%%%%%%%%%%%%%%%%%%%%%%%%%%%%%%%
%%%%%%%%%%%%%%%%%%%%%%%%%%%%%%%%%%%%%%%%%%%%%%

%%%%%%%%%%%%%%%%%%%%%%%%%%%%%%%%%%%%%%%%%%%%%%
%%%%%%%%%%%%%%%%%%%%%%%%%%%%%%%%%%%%%%%%%%%%%%
\begin{figure}
\begin{centering}
\includegraphics[width=0.5\textwidth]{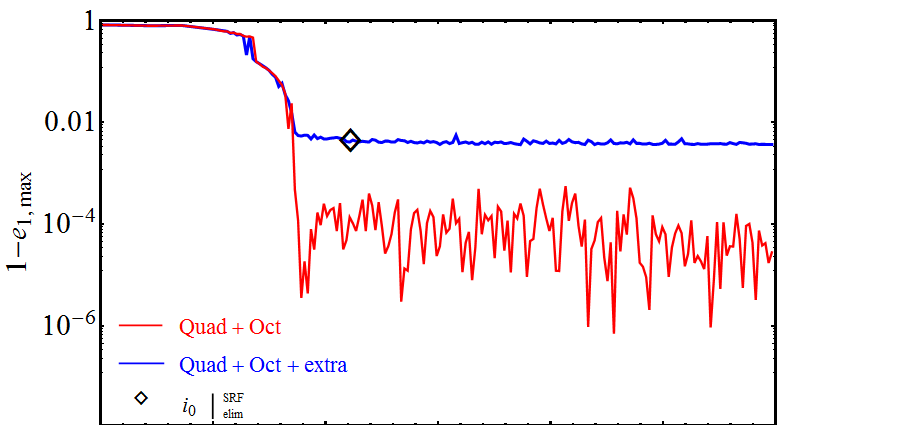}
\includegraphics[width=0.5\textwidth]{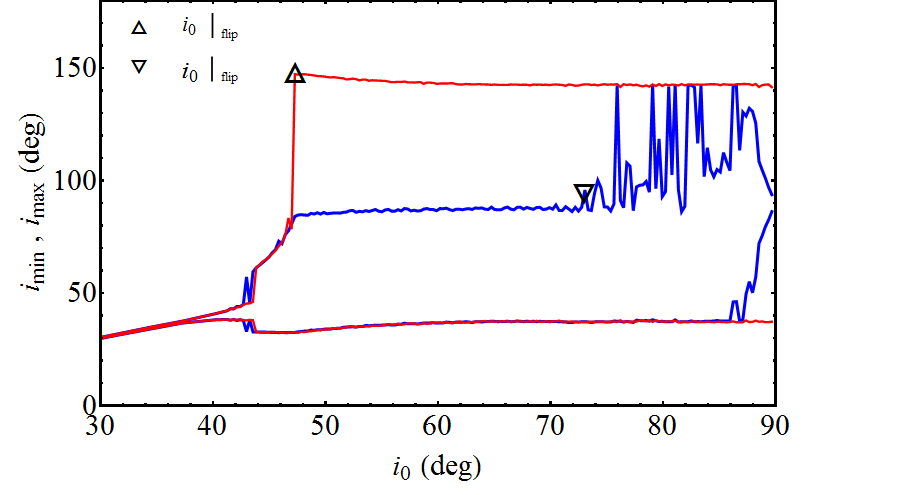}
\caption{Case 10b in Table 1 ($\varepsilon_{\oct}=0.067$,
$\varepsilon_\gr=2.37\times10^{-4}$, $\varepsilon_\tide=1.39\times10^{-9}$).
System parameters are $m_0=1M_\odot$, $m_1=1M_J$, $m_2=1M_\odot$
 $a_1=1\au$ and $a_2=33.33\au$. Orbits are started with $e_1=0.001$, $e_2=0.8$,
 {$\omega_1=0^\circ$, $\omega_2=0^\circ$, $\Omega_1=0^\circ$ and $\Omega_2=180^\circ$.}
The total integration time is $5\times10^{5}$ years ($\sim 392.5~t_K$).
Note that $e_\li$ decreases,
the critical angles ($i_0|_{e\li}^{\mathrm{SRF}}$ and $i_0|_{\mathrm{flip}}^{\mathrm{SRF}}$) are pushed to higher values.
}
\label{fig:case10b}
\end{centering}
\end{figure}
%%%%%%%%%%%%%%%%%%%%%%%%%%%%%%%%%%%%%%%%%%%%%%
%%%%%%%%%%%%%%%%%%%%%%%%%%%%%%%%%%%%%%%%%%%%%%

%%%%%%%%%%%%%%%%%%%%%%%%%%%%%%%%%%%%%%%%%%%%%%
%%%%%%%%%%%%%%%%%%%%%%%%%%%%%%%%%%%%%%%%%%%%%%
\begin{figure}
\begin{centering}
\includegraphics[width=0.5\textwidth]{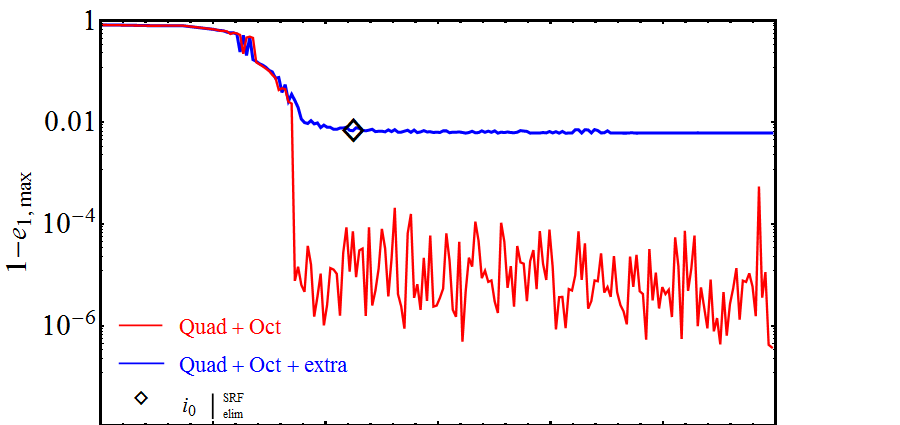}
\includegraphics[width=0.5\textwidth]{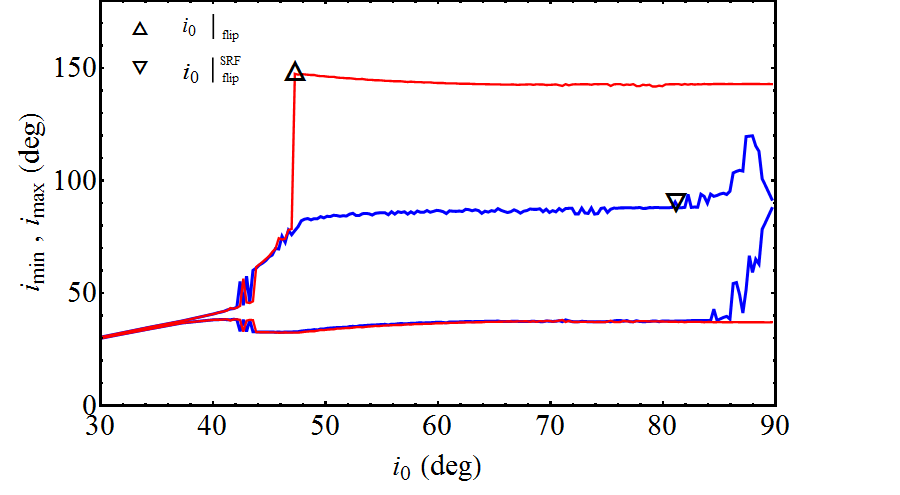}
\caption{Case 10c in Table 1 ($\varepsilon_{\oct}=0.067$,
$\varepsilon_\gr=2.37\times10^{-3}$, $\varepsilon_\tide=1.39\times10^{-8}$).
System parameters are $m_0=1M_\odot$, $m_1=1M_J$, $m_2=0.1M_\odot$
 $a_1=1\au$ and $a_2=33.33\au$. Orbits are started with $e_1=0.001$, $e_2=0.8$,
 {$\omega_1=0^\circ$, $\omega_2=0^\circ$, $\Omega_1=0^\circ$ and $\Omega_2=180^\circ$.}
The total integration time is $5\times10^{6}$ years ($\sim 392.5~t_K$).
The value of $e_\li$ becomes smaller than in Case 10b and
the maximum inclination angle cannot reach $140^\circ$.
}
\label{fig:case10c}
\end{centering}
\end{figure}
%%%%%%%%%%%%%%%%%%%%%%%%%%%%%%%%%%%%%%%%%%%%%%
%%%%%%%%%%%%%%%%%%%%%%%%%%%%%%%%%%%%%%%%%%%%%%

%%%%%%%%%%%%%%%%%%%%%%%%%%%%%%%
\subsection{Parameter space}
We have carried out calculations of the inner binary evolution for
various combinations of $a_1,a_2,m_2,e_2$ that yield different values
of the parameters $\varepsilon_\oct$, $\varepsilon_\gr$, $\varepsilon_\tide$, $\varepsilon_\rot$ (see Table 1).  
In particular, the
dimensionless octupole parameter $\varepsilon_\oct$ can be varied by
changing the values of $e_2$ and the ratio of $a_1$ to $a_2$ (see
  Equation~\ref{eq:C}), and we consider $\varepsilon_\oct$ ranging from
$0.001$ to $0.1$. For a given $\varepsilon_\oct$,
we consider various possible values of $\varepsilon_\gr$, $\varepsilon_\tide$, $\varepsilon_\rot$
(see Equations~\ref{eq:R11}, \ref{eq:R12} and \ref{eq:R13}) in order to
assess the role of SRFs.
For each set of parameters and the initial inclination angle $i_0$,
we integrate the binary evolution equations for a few
octupole oscillation periods, $t_K/\varepsilon_\oct~$,
and record the maximum of $e_1$ and the extrema of $\I$ attained during the
evolution. Table 2 summarizes our key findings.

As noted before (Section 4.1), the SRFs provide an upper limit to the
maximum eccentricity attainable during the binary evolution, even for
large $\varepsilon_\oct$. In particular, our numerical result
for the maximum eccentricity $e_{\rm max,Num}$ (for all $i_0$'s)
is in good agreement with the limiting eccentricity $e_{\rm lim}$
given by Equation~(\ref{eq:FTT}). Comparing
$\Dot\omega_\extra/\Dot\omega_\mathrm{K}$ at $e_1=e_{\rm lim}$, we see that
with the exception of Case 1, the tidal effect and the rotational bulge
are responsible for limiting the eccentricity growth.

The last three columns of Table 2 summarize the three critical initial
inclination angles introduced in Section 4.1 for the different cases.
Without SRFs, the angle $i_0|_{\mathrm{flip}}$ (at which orbital flip
occurs due to the octupole potential) decreases with increasing
$\varepsilon_\oct$.  When the SRFs are included, orbital flips require
higher inclinations
($i_0|_{\mathrm{flip}}^{\mathrm{SRF}}>i_0|_{\mathrm{flip}}$), and the
critical angle $i_0|_{\mathrm{flip}}^{\mathrm{SRF}}$ decreases as
$\varepsilon_\oct$ increases. Note that when $\varepsilon_\oct\lesssim
0.02$ (Case 1 to Case 6), $i_0|_{\mathrm{flip}}^{\mathrm{SRF}}\approx
90^\circ$, implying that the octupole potential cannot lead to orbit
flip. Finally, in the presence of the SRFs, the critical inclination
$i_0|_{e\li}^{\mathrm{SRF}}$ at which the maximum eccentricity
saturates to $e_{\lim}$ is roughly equals to $i_0|_{\mathrm{flip}}$.
This implies that the excitations of eccentricity and inclination are
related.

Figures \ref{fig:case10a}\---\ref{fig:case10c} depict the results for Case 10a-10c,
corresponding to the same $\varepsilon_\oct$ but different
SRF strength. Note that, as $\varepsilon_\oct$ is the same for all these examples,
the width of the octupole window of influence is unaltered. However, the
region in inclination angle for which orbits are allowed to flip changes with
$\varepsilon_\mathrm{extra}$. This is quantified by the value
of the critical angle $i_0|_{\mathrm{flip}}^{\mathrm{SRF}}$, which grows
monotonically with $\varepsilon_\mathrm{extra}$, meaning that orbital
flips are progressively confined to the neighboring region of $i_0=90^\circ$.

%%%%%%%%%%%%%%%%%%%%%%%%%%%%%%%%%%%%%%%%%%%%%%
\section{Numerical experiments in the comparable mass regime}
In this section, we extend the analysis of previous sections to the general case of
Lidov\---Kozai cycles with SRFs
in systems composed of three
comparable masses ($m_0\sim m_1\sim m_2\sim m_\odot$), focusing on the long
term evolution of eccentricity and inclination of the inner binary.

%%%%%%%%%%%%%%%%%%%%%%%%%%%%%%%%%
\subsection{Symmetry in inclination at the quadrupole-level approximation}\label{sec:symmetry_general}
As discussed in Section~\ref{sec:symmetry}, the equations
of motion in the test-particle limit are symmetric upon reflections of the $\jvec_1$ vector
through the origin. This implies that the Lidov\---Kozai cycles with SRFs examples in the
small mass regime ($m_1\ll m_0$) presented in Figures~\ref{fig:case94}\---\ref{fig:case10c}
show even symmetry around $i_0=90^\circ$ in the eccentricity curves (top panels) and
odd symmetry in the inclination curves (bottom panels).
As we have shown in a previous example (Figure~\ref{fig:SCM}), this reflection symmetry 
is removed when $m_1\sim m_0$.

However, there is still an approximate symmetry center for calculations
at the quadrupole-level. This can be seen in Figure~\ref{fig:case1c} for a triple system of comparable masses. In a similar fashion
to Figures~\ref{fig:case94}\---\ref{fig:case10c}, the black curves in Figure~\ref{fig:case1c} show $e_{1,\m}$ and
$i_{1,\mathrm{max/min}}$ as a function of $i_{1,0}$ and $i_{\mathrm{tot},0}$ calculated from the quadrupole-level
potential. We have extended the initial inclinations to cover $(0^\circ$,$180^\circ)$, encompassing
the full range of prograde and retrograde orientations.
There is a reflection symmetry respect to
$i_{\mathrm{tot},0}\approx94.5^\circ$ (or equivalently, respect to $i_{1,0}\approx85.5^\circ$).
However, this symmetry is erased once octupole-level terms are considered (red curves). This
is in contrast to the test-particle limit, for which the reflection symmetry around
$i_{\mathrm{tot},0}=i_{1,0}=90^\circ$ is valid for the quadrupole-level and octupole-level approximations.

%%%%%%%%%%%%%%%%%%%%%%%%%%%%%%%%%%%%%%%%%%%%%%%%%%
%%%%%%%%%%%%%%%%%%%%%%%%%%%%%%%%%%%%%%%%%%%%%%%%%%
\begin{table*}
 \centering
 \begin{minipage}{150mm}
  \caption{Initial conditions on different cases: $m_0=1M_\odot$, $e_0=0.001$, $k_{2,1}=0.014$,
  {$\omega_1=\omega_2=\Omega_1=0^\circ$ and $\Omega_2=180^\circ$.}
The parameter  $\varepsilon_\extra$ is calculated from the definition in Equations~(\ref{eq:R11}) and (\ref{eq:R12}).
   \label{tab:paramspace_general}}
  \begin{tabular}{@{}llrrrrlrlr@{}}
  \hline
   Parameter & $\varepsilon_{\oct}$ & $\varepsilon_{\gr}$ ~~~~~& $\varepsilon_{\tide}$~~~~~
     & $a_1 (\au)$ & $a_2(\au)$ & $e_{2,0}$~ & $m_1(M_\odot)$ & $m_2(M_\odot)$~~~~~ & $R_1(R_\odot)$  \\
 \hline
 Case 1a & 0.022 & $4.33\times10^{-6}$ & $4.00\times10^{-16}$ & 10~~~~ & 100~~~ & ~0.5 & 0.5~~~~~~& ~~~~1 & 0.5~~~~~ \\
 Case 1b & 0.022 & $1.44\times10^{-5}$ & $1.64\times10^{-13}$  & 3~~~~~ & 30~~~~ & ~0.5 & 0.5~~~~~  & ~~~~1 & 0.5~~~~~ \\
 Case 1c & 0.022 & $4.33\times10^{-5}$ & $4.00\times10^{-11}$  & 1~~~~~ & 10~~~~ & ~0.5 & 0.5~~~~~  & ~~~~1 & 0.5~~~~~ \\
 Case 1d & 0.022 & $4.33\times10^{-5}$ & $4.10\times10^{-8}$~  & 1~~~~~ & 10~~~~ & ~0.5 & 0.5~~~~~  & ~~~~1 & 2~~~~~~ \\
 Case 1e & 0.022 & $4.33\times10^{-5}$ & $4.00\times10^{-6}$~  & 1~~~~~ & 10~~~~ & ~0.5 & 0.5~~~~~  & ~~~~1 & 5~~~~~~ \\
 \\
 Case 2a & 0.042 & $5.53\times10^{-6}$ & $7.64\times10^{-17}$ & 10~~~~ & 120~~~ & ~0.6 & 0.3~~~~~~& ~~~0.8 & 0.3~~~~~ \\
 Case 2b & 0.042 & $1.11\times10^{-5}$ & $2.45\times10^{-15}$  & 5~~~~~ & 60~~~~ & ~0.6 & 0.3~~~~~  & ~~~0.8 & 0.3~~~~~ \\
 Case 2c & 0.042 & $5.53\times10^{-5}$ & $7.64\times10^{-12}$  & 1~~~~~ & 12~~~~ & ~0.6 & 0.3~~~~~  & ~~~0.8 & 0.3~~~~~ \\
 Case 2d & 0.042 & $5.53\times10^{-5}$ & $2.45\times10^{-10}$  & 1~~~~~ & 12~~~~ & ~0.6 & 0.3~~~~~  & ~~~0.8 & 0.6~~~~~ \\
 Case 2e & 0.042 & $5.53\times10^{-5}$ & $9.83\times10^{-6}$~  & 1~~~~~ & 12~~~~ & ~0.6 & 0.3~~~~~  & ~~~0.8 & 5~~~~~~ \\
\hline
\end{tabular}
\end{minipage}
\end{table*}
%%%%%%%%%%%%%%%%%%%%%%%%%%%%%%%%%%%%%%%%%%%%%%%%%%
%%%%%%%%%%%%%%%%%%%%%%%%%%%%%%%%%%%%%%%%%%%%%%%%%%

%%%%%%%%%%%%%%%%%%%%%%%%%%%%%%%%%%%%%%%%%%%%%%%%%%
%%%%%%%%%%%%%%%%%%%%%%%%%%%%%%%%%%%%%%%%%%%%%%%%%%
\begin{table*}
 \centering
 \begin{minipage}{140mm}
  \caption{Results on different cases. $i_{1,0}|_{\mathrm{sym}}$ is the initial inclination where $e_{1,\m}$
  reaches the maximum at the quadrupole level. While $i_{\mathrm{tot},0}|_{\mathrm{sym}}$ is for total initial angle.
\label{tab:results_general}
          }
  \begin{tabular}{@{}llrrrrlrlr@{}}
\hline
   Parameter &  ~~~$1-e_\li$ & $1-e_{\m,{\mathrm{Num}}}$  & $\left.\frac{\Dot \omega_\gr}{\Dot \omega_\mathrm{K}}\right|_{e_\li} $
      & $\left.\frac{\Dot \omega_\tide}{\Dot \omega_\mathrm{K}}\right|_{e_\li}$
      & $i_{1,0}|_{\mathrm{sym}}$(deg)
      & $i_{\mathrm{tot},0}|_{\mathrm{sym}}$(deg)\\
 \hline
 Case 1a & $1.40\times10^{-4}$ & $1.39\times10^{-4}$ ~  & ~~~$2.58\times10^{-4}$  & 10.12~~~~~ & 85.5~~~~~~   & ~~~~~~94.5 \\
 Case 1b & $5.35\times10^{-4}$ & $5.27\times10^{-4}$ ~  & ~~~$4.43\times10^{-4}$  & 10.11~~~~~ & 85.5~~~~~~   & ~~~~~~94.5\\
 Case 1c & $1.81\times10^{-3}$ & $1.80\times10^{-3}$ ~  & ~~~$7.19\times10^{-4}$  & 10.09~~~~~ & 85.5~~~~~~   & ~~~~~~94.5\\
 Case 1d & $8.50\times10^{-3}$ &  $8.46\times10^{-3}$ ~  & ~~~$3.33\times10^{-4}$  & 9.98~~~~~~ & 85.5~~~~~~   & ~~~~~~94.5   \\
 Case 1e & $2.37\times10^{-2}$ &  $2.36\times10^{-2}$ ~  & ~~~$2.00\times10^{-4}$  & 9.74~~~~~~ & 85.5~~~~~~   & ~~~~~~94.5   \\
 \\
 Case 2a & $9.71\times10^{-5}$ &  $9.57\times10^{-5}$ ~  & ~~~$3.97\times10^{-4}$  & 10.12~~~~~ &  86.2~~~~~~  & ~~~~~~93.8\\
 Case 2b & $2.10\times10^{-4}$ &  $2.06\times10^{-4}$ ~  & ~~~$5.40\times10^{-4}$  & 10.12~~~~~ &  86.2~~~~~~  & ~~~~~~93.8 \\
 Case 2c & $1.26\times10^{-3}$ &  $1.22\times10^{-3}$ ~  & ~~~$1.10\times10^{-3}$  & 10.10~~~~~ &  86.2~~~~~~  & ~~~~~~93.8  \\
 Case 2d & $2.71\times10^{-3}$ &  $2.64\times10^{-3}$ ~  & ~~~$7.52\times10^{-4}$  & 10.07~~~~~ &  86.2~~~~~~  & ~~~~~~93.8  \\
 Case 2e & $2.90\times10^{-2}$ &  $2.89\times10^{-2}$ ~  & ~~~$2.31\times10^{-4}$  & 9.65~~~~~~ &  86.2~~~~~~  & ~~~~~~93.8  \\
\hline
\end{tabular}
\end{minipage}
\end{table*}
%%%%%%%%%%%%%%%%%%%%%%%%%%%%%%%%%%%%%%%%%%%%%%%%%%
%%%%%%%%%%%%%%%%%%%%%%%%%%%%%%%%%%%%%%%%%%%%%%%%%%

The shift in the symmetry center away from $90^\circ$ results
from the conservation of total angular momentum and the quadrupole-level
potential. Introducing the total angular momentum $G_{\mathrm{tot}}$, we can
write the mutual inclination of the inner and outer orbits as \citep{Smadar 2013b}
\begin{equation}\label{eq:GTOT}
\cos\I=\frac{G_{\mathrm{tot}}^2-L_1^2(1-\e1)-L_2^2(1-e_2^2)}{2L_1L_2\sqrt{1-\e1}\sqrt{1-e_2^2}}~,
\end{equation}
where $L_1$ and $L_2$ are given by Equations (\ref{eq:L1})\---(\ref{eq:L2}).
Since $G_\mathrm{tot}$ is conserved, this expression becomes
\begin{equation}\label{eq:GTOT2}
\cos\I=\frac{\cos i_{\mathrm{tot},0}}{\sqrt{1-\e1}}+\frac{L_1}{2L_2}\frac{\e1}{\sqrt{{1-\e1}\vphantom{1-e_{2,0}^2}}\sqrt{1-e_{2,0}^2}}~,
\end{equation}
where $i_{\mathrm{tot},0}$, $e_{1,0}=0$ and $e_{2,0}$ are the initial values
for the mutual inclination and for the inner and outer eccentricities, respectively.
Note that, at the quadrupole level, the eccentricity of the outer orbit
$e_2=e_{2,0}=$~constant. Next, we rewrite the (constant) potential of Equation~(\ref{eq:APK2})
as
\begin{equation}\label{eq:APK3}
\begin{split}
-\frac{8\langle\Phi_\K\rangle}{\mu_1\Phi_0}&=3\cos^2\!\I(1+4\e1)-1-9\e1\\
&+15\e1\cos^2\!\omega_1(1-\cos^2\!\I)~.
\end{split}
\end{equation}
Combining Equations (\ref{eq:GTOT2}) and (\ref{eq:APK3}), we obtain an
expression for the maximum eccentricity $e_{1,\m}$ (after evaluating at $\omega_1=\pi/2$):
\begin{equation}\label{eq:MEG}
\begin{split}
&5\cos^2\!i_{\mathrm{tot},0}-3+\frac{L_1}{L_2}\frac{\cos i_{\mathrm{tot},0}}{\sqrt{1-e_{2,0}^2}}+\Big(\frac{L_1}{L_2}\Big)^2\frac{e_{1,\m}^4}{1-e_{2,0}^2}\\
&+e_{1,\m}^2\Bigg[3+4\frac{L_1}{L_2}\frac{\cos i_{\mathrm{tot},0}}{\sqrt{1-e_{2,0}^2}}+\Big(\frac{L_1}{2L_2}\Big)^2\frac{1}{1-e_{2,0}^2}
\Bigg]=0~.
\end{split}
\end{equation}
Equation~(\ref{eq:MEG}) generalizes Equation~(\ref{eq:ME}) for the test-mass case.
These two expressions become equivalent in the limit $L_1/L_2\rightarrow0$.

In the test-particle limit at the quadrupole level, the symmetry center (at $i_0=90^\circ$)
coincides with the point of maximal ``eccentricity" or $e_{1,\m}=1$.
Similarly, for the general case of comparable masses,
we define the angle $i_{\mathrm{tot},0}|_{\mathrm{sym}}$ by setting $e_{1,\m}=1$ in
Equation~(\ref{eq:MEG}):
\begin{equation}\label{eq:isym}
\cos i_{\mathrm{tot},0}|_{\mathrm{sym}}=-\frac{1}{2}\frac{L_1}{L_2}\frac{1}{\sqrt{1-e^2_{2,0}}}~.
\end{equation}
As expected, in the limit $L_1/L_2\rightarrow0$ we have that
 $i_{\mathrm{tot},0}|_{\mathrm{sym}}\rightarrow90^\circ$.
Note that the negative sign in Equation~(\ref{eq:isym}) implies
 that $i_{\mathrm{tot},0}|_{\mathrm{sym}}$ is always greater than $90^\circ$.
This symmetry breaking can also be realized for the inner inclination $i_1$, which can
be obtained from conservation of angular momentum and the law of sines:
$L_1(1-e^2_{1,0})^{1/2}/\sin\!i_{2,0}=G_{\mathrm{tot}}/\sin\!i_{\mathrm{tot},0}=L_2(1-e^2_{2,0})^{1/2}/\sin\!i_{1,0}$.
This is consistent with the numerical result shown in Figure~\ref{fig:case1c}.

%%%%%%%%%%%%%%%%%%%%%%%%%%%%%%%%%%%%%%
\subsection{Parameter space}\label{sec:paramspace_general}
\begin{figure}
\begin{centering}
\includegraphics[width=8.5cm]{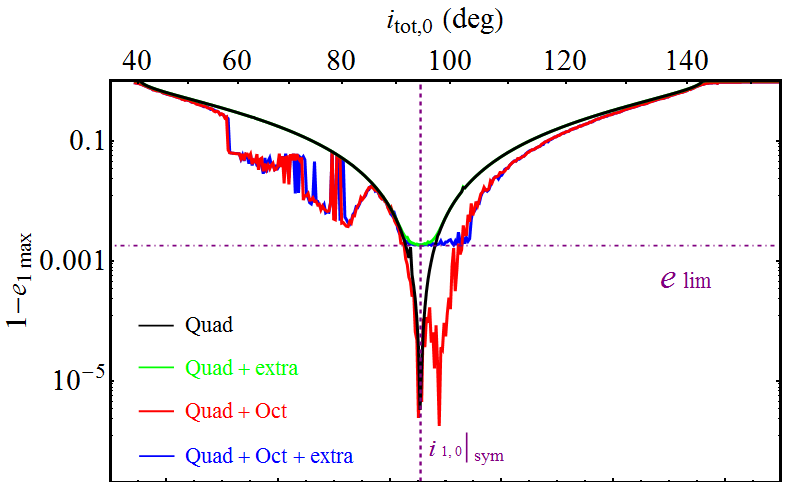}
\includegraphics[width=8.5cm]{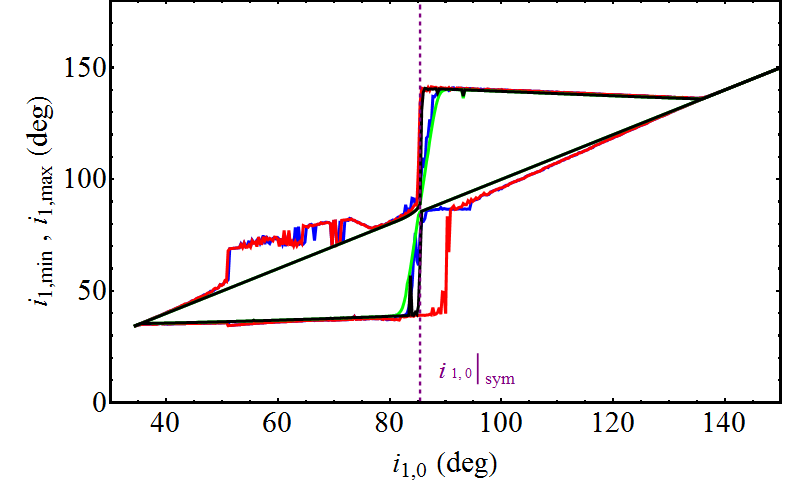}
\caption{CASE 1c of Table 3. $\varepsilon_{\oct}=0.022$.
We extend the test-particle case in Section 4 to comparable-mass case.
Here,
the system has an inner binary of $m_0=1M_\odot$, $m_1=0.5M_\odot$, $a_1=1\au$, $R_1=0.5R_\odot$
and the companion has $m_2=1M_\odot$, $a_2=10\au$.
We initially set $e_1=0.001$, $e_2=0.5$, {$\omega_1=0^\circ$, $\omega_2=0^\circ$, $\Omega_1=0^\circ$ and $\Omega_2=180^\circ$.}
Because there
is no reflection symmetry between the prograde
and retrograde configurations,
we integrate Equations (\ref{eq:fulle1})-(\ref{eq:fullomega2}) with $i_{1,0}\in(30^\circ,150^\circ)$ for $5\times10^4$ years ($\sim395.0t_K$).
Note that limiting eccentricity holds as well and the flip cannot occur with SRFs.
}
\label{fig:case1c}
\end{centering}
\end{figure}

Figures~\ref{fig:case1c}\---\ref{fig:case2e} show the dependence of maximum eccentricity
and inclination extrema as a function of $i_{\mathrm{1},0}$ (or $i_{\mathrm{tot},0}$)
for systems with different masses $m_0=1M_\odot$, $m_1=0.5M_\odot$
and other parameters. As in the examples of Section~\ref{sec:paramspace}, four different calculations shown:
(1) quadrupole-level approximation with no SRFs (black curves),
(2) octupole-level approximation with no SRFs (red curves), (3) quadrupole-level
approximation with SRFs (green curves), and (4) octupole-level approximation with SRFs (blue curves).

Just as in Section~\ref{sec:paramspace}, we explore the changes in $e_{1,\m}$ and
$i_{1,\mathrm{max/min}}$ as we vary $\varepsilon_\oct$ and $\varepsilon_\mathrm{extra}$, this
time for triple stars of comparable masses.
Note that in Equation~(\ref{eq:C}), the octupole contribution is exactly zero if the members of inner binary here equal mass,
so for similar masses, $\varepsilon_\oct$ cannot be very large. The orbital separations and other physical
parameters of the systems studied in this section are listed in Table~\ref{tab:paramspace_general}.

The values of initial inclinations induced both prograde and retrograde orbits in $i_{1,0}$ (i.e., respect to the
total angular momentum vector). The
range in angles is chosen so as to enclose the ``Lidov\---Kozai active" region. This region is contained between
the angles  $i_{\mathrm{tot},0}|_{\mathrm{st}}^-$ and $i_{\mathrm{tot},0}|_{\mathrm{st}}^+$, which are the two
solutions of the quadratic equation obtained from Equation~(\ref{eq:MEG}) after setting $e_{1,\m}=0$:
\begin{equation}\label{eq:ist}
5\cos^2i_{\mathrm{tot},0}|_{\mathrm{st}}-3+\frac{L_1}{L_2}\frac{\cos i_{\mathrm{tot},0}|_{\mathrm{st}}}{\sqrt{1-e^2_{2,0}}}=0~.
\end{equation}
For all the examples considered here (Table~\ref{tab:paramspace_general}), an inclination interval of
 $i_{1,0}\in(30^\circ,150^\circ)$ is sufficient to capture the entire range of systems that are subject
 to Lidov\---Kozai oscillations.

%%%%%%%%%%%%%%%%%%%%%%%%%%%%%%%%%%%%%%%%%%%%%%
%%%%%%%%%%%%%%%%%%%%%%%%%%%%%%%%%%%%%%%%%%%%%%
%%%%%%%%%%%%%%%%%%%%%%%%%%%%%%%%%%%%%%%%%%%%%%
%%%%%%%%%%%%%%%%%%%%%%%%%%%%%%%%%%%%%%%%%%%%%%

%%%%%%%%%%%%%%%%%%%%%%%%%%%%%%%%%%%%%%%%%%%%%%
%%%%%%%%%%%%%%%%%%%%%%%%%%%%%%%%%%%%%%%%%%%%%%
\begin{figure}
\begin{centering}
\includegraphics[width=8.5cm]{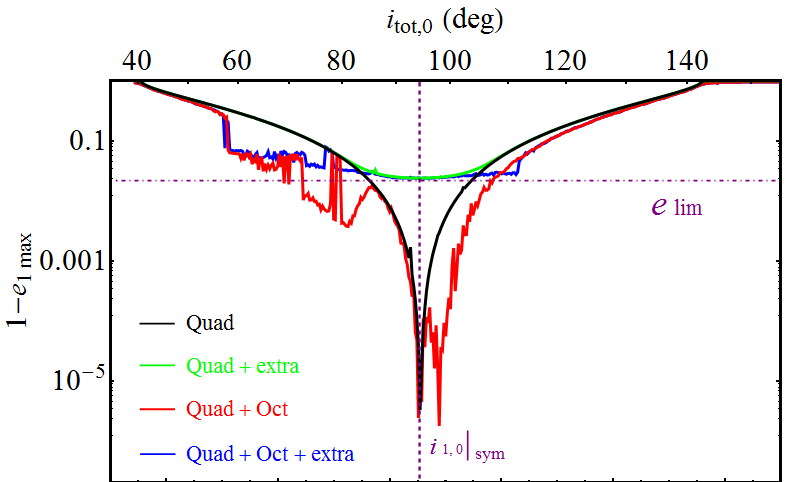}
\includegraphics[width=8.5cm]{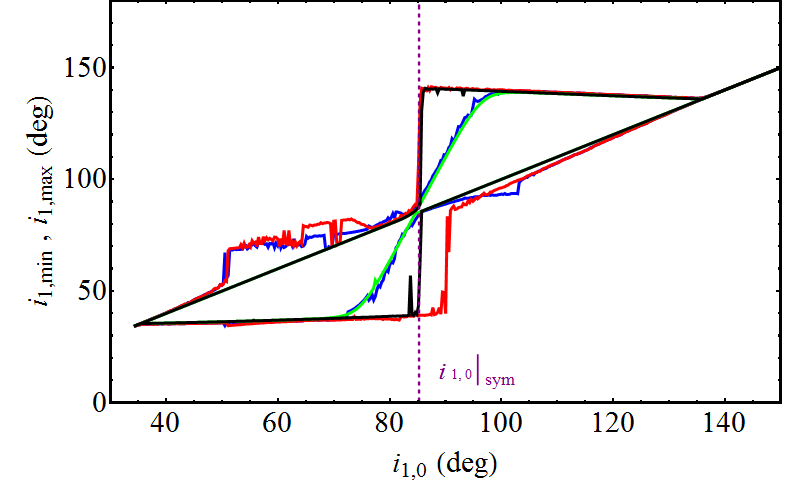}
\caption{CASE 1e of Table 3.
We increase $\varepsilon_{\extra}$ by changing $a_1=1\au$, $a_2=10\au$ and $R_1=5R_\odot$,
and keep other quantities the same as Case 1c (See Table 3).
We integrate the equations and the total integration time is $5\times10^4$ years ($\sim395.0t_K$).
Note that $e_\li$ become smaller and SRFs affect the octupole-level effects significantly.
}
\label{fig:case1e}
\end{centering}
\end{figure}
%%%%%%%%%%%%%%%%%%%%%%%%%%%%%%%%%%%%%%%%%%%%%%
%%%%%%%%%%%%%%%%%%%%%%%%%%%%%%%%%%%%%%%%%%%%%%

%%%%%%%%%%%%%%%%%%%%%%%%%%%%%%%%%%%%%%%%%%%%%%
%%%%%%%%%%%%%%%%%%%%%%%%%%%%%%%%%%%%%%%%%%%%%%
\begin{figure}
\begin{centering}
\includegraphics[width=8.5cm]{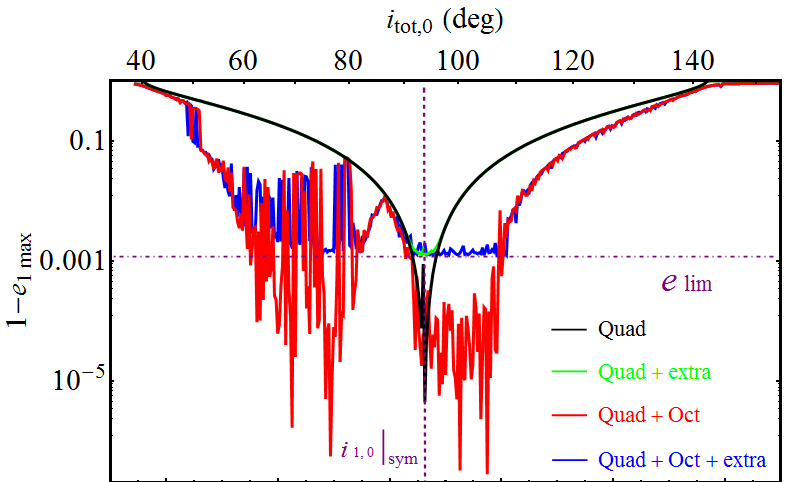}
\includegraphics[width=8.5cm]{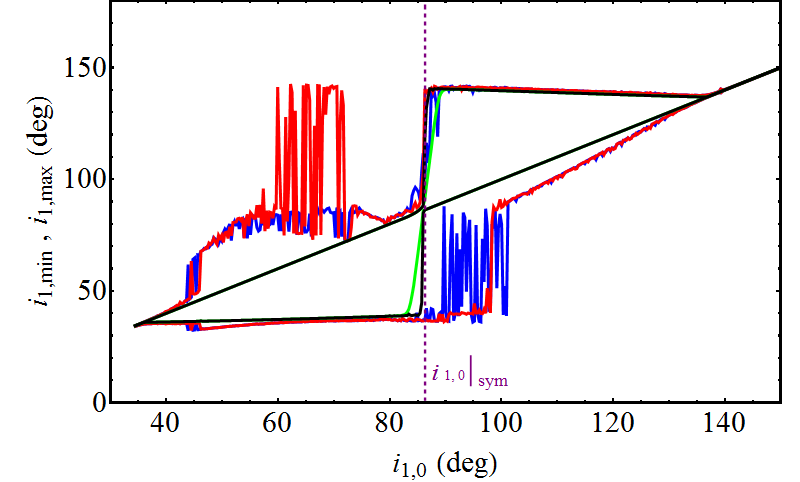}
\caption{CASE 2c of Table 3.
We carry out another set of
numerical integrations with higher $\varepsilon_{\oct}=0.042$.
The system has an inner binary of $m_0=1M_\odot$, $m_1=0.3M_\odot$, $a_1=1\au$, $R_1=0.3R_\odot$
and the companion has $m_2=0.8M_\odot$, $a_2=12\au$.
We initially set $e_1=0.001$, $e_2=0.6$,{$\omega_1=0^\circ$, $\omega_2=0^\circ$, $\Omega_1=0^\circ$ and $\Omega_2=180^\circ$.}
We integrate the equations for $1.2\times10^5$ years ($\sim598.0t_K$).
Due to the stronger octupole potential,
the width of the $e_{1,\m}$ and $i_{0,\m/\mi}$ regions become lager
and the flip occur even with SRFs.
}
\label{fig:case2c}
\end{centering}
\end{figure}
%%%%%%%%%%%%%%%%%%%%%%%%%%%%%%%%%%%%%%%%%%%%%%
%%%%%%%%%%%%%%%%%%%%%%%%%%%%%%%%%%%%%%%%%%%%%%

%%%%%%%%%%%%%%%%%%%%%%%%%%%%%%%%%%%%%%%%%%%%%%
%%%%%%%%%%%%%%%%%%%%%%%%%%%%%%%%%%%%%%%%%%%%%%
\begin{figure}
\begin{centering}
\includegraphics[width=8.5cm]{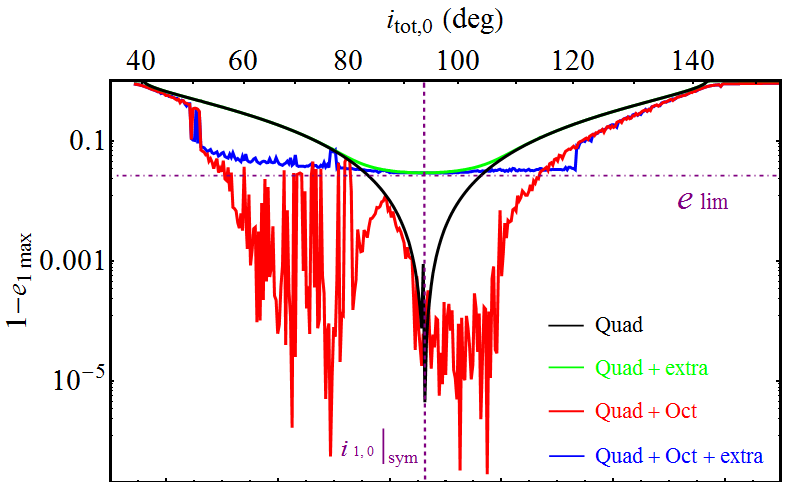}
\includegraphics[width=8.5cm]{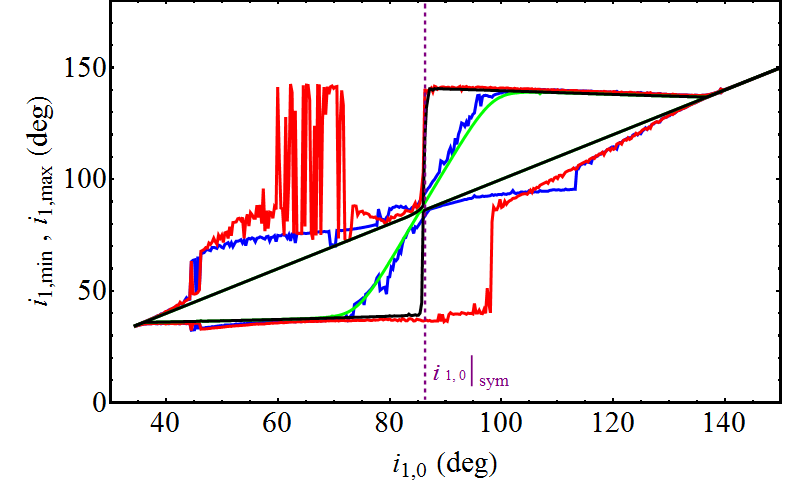}
\caption{CASE 2e of Table 3.
Results of numerical integrations with higher $\varepsilon_\extra$.
The system has an inner binary of $m_0=1M_\odot$, $m_1=0.3M_\odot$, $a_1=1\au$, $R_1=5R_\odot$
and the companion has $m_2=0.8M_\odot$, $a_2=12\au$.
We initially set $e_1=0.001$, $e_2=0.6$, {$\omega_1=0^\circ$, $\omega_2=0^\circ$, $\Omega_1=0^\circ$ and $\Omega_2=180^\circ$.}
The total  integration time is $1.2\times10^5$ years ($\sim598.0t_K$).
}
\label{fig:case2e}
\end{centering}
\end{figure}
%%%%%%%%%%%%%%%%%%%%%%%%%%%%%%%%%%%%%%%%%%%%%%
%%%%%%%%%%%%%%%%%%%%%%%%%%%%%%%%%%%%%%%%%%%%%%

%%%%%%%%%%%%%%%%%%%
\subsubsection{Eccentricity maxima}
As discussed in Section~\ref{sec:symmetry_general} above, for finite $m_1$, the symmetry
of the system respect to $i_{\mathrm{tot},0}=90^\circ$ is shifted to a different value
$i_{\mathrm{tot},0}|_{\mathrm{sym}}$. However, this symmetry center only reflects
the behaviour of the system at the quadrupole level. In addition to the significant when the octupole potential is included,
Figure~\ref{fig:case1c} (for $\varepsilon_\oct=0.022$) also shows how the symmetry between prograde and retrograde
orbits is broken.

Figure \ref{fig:case1c} shows that at the quadrupole level,
the excitation of
eccentricity only takes place in the range between
$i_{\mathrm{tot},0}|_{\mathrm{st}}=40.4^\circ$
and $i_{\mathrm{tot},0}|_{\mathrm{st}}=142.0^\circ$, as given
by Equation~(\ref{eq:ist}),
and $e_{1,\m}$ is achieved at $i_{\mathrm{tot},0}|_{\mathrm{sym}}\approx85.5^\circ$
(Table 4), as predicted by Equation (\ref{eq:isym}).
When the octupole-level terms are included
(red curve), $i_{\mathrm{tot},0}\approx85.5^\circ$
is not the only inclination that allows for such extreme eccentricity (i.e., $e_{1,\m}\approx1$).
Indeed, as in the test-mass case (Figures~\ref{fig:case94}--\ref{fig:case10c}),
the inclusion of octupole terms widens the range on angles for which $e_1\rightarrow 1$ is
possible. In this case however, the ``widening" takes place to the right of
$i_{\mathrm{tot},0}|_{\mathrm{sym}}$, while the quadrupole-level
solution remains valid for $i_{1,0}<i_{\mathrm{tot},0}|_{\mathrm{sym}}$ down to for $i_{1,0}\sim75^\circ$.
Below $\sim75^\circ$, deviations from the quadrupole-level solution become significant.
These features, which can be compared to what seemed to be minor fluctuations in $e_{1,\m}$ for the test-particle cases
(see Figures~\ref{fig:case94}--\ref{fig:case10c}), show that moderately high eccentricities can be excited
at lower inclinations ($i_{\mathrm{tot},0}=60^\circ\sim80^\circ$) than the quadrupole-order calculation would allow.

When SRFs are included,
the quadrupole-level eccentricity maxima are truncated at
a global maximum corresponding to $1-e_{1,\m}\sim10^{-3}$.
The horizontal line in Figure \ref{fig:case1c} corresponds to the
limiting eccentricity given by Equation~(\ref{eq:FTT}).
It is important to note that this limiting
eccentricity applies even in the general
case of comparable masses.
Also note that the tides are mainly responsible for the eccentricity suppression
(see Table 4).

It is not surprising that the analytic estimate of $e_\mathrm{lim}$,
derived in the test-mass limit (Section~\ref{sec:max_ecc}),
remains a good approximation for comparable-mass systems.
For $e_1$ very close to unity,
the vast majority of the angular momentum resides in the outer binary, forcing the inner
binary to behave essentially as a test particle. Note that we can use
Equations~(\ref{eq:j1vec}),~(\ref{eq:j2vec}) and ~(\ref{eq:e2vec})
at the quadrupole level ($\varepsilon_\oct=0$) to write
\begin{equation}
\begin{split}
\frac{d}{dt}\big({\jvec_1\cdot\nvec_2}\big)&= \frac{1}{\sqrt{1-e_2^2}}~\jvec_1\cdot\frac{d\jvec_2}{dt}\\
&=\frac{15}{4t_K}\frac{L_1\sqrt{1-e_1^2}}{L_2\sqrt{1-e_2^2}} e_1^2(\uvec_1\cdot\nvec_2)(~\uvec_1\times\nvec_2)\cdot \nvec_1~~,
\end{split}
\end{equation}
where all the vectors involved are of norm unity.  Then we find that
$d\big({\jvec_1\cdot\nvec_2}\big)/dt$ is very small provided that
$L_2\sqrt{1-e_2^2}\gg L_1\sqrt{1-e_1^2}$. Therefore, for the high eccentricity
phase of Lidov\---Kozai cycles, the analysis
of Section~\ref{sec:max_ecc} for test particles still applies in the comparable-mass regime.
As in the test-mass cases (Section~\ref{sec:paramspace}),
the octupole effect expands the range of the initial mutual
inclinations capable of reaching maximal eccentricities.

%%%%%%%%%%%%%%%%%%%
\subsubsection{Inclination extrema}
As for the eccentricity curves, the inclination curves at the quadrupole level
in Figure~\ref{fig:case1c} (black curves) are symmetric respect to $i_{1,0}|_{\mathrm{sym}}$.
In this particular example, SRFs do not inflict
significant modifications except for the close vicinity of $i_{1,0}|_{\mathrm{sym}}$, which sees
the amplitude in the inclination oscillations (the difference between $i_{1,\mathrm{min}}$ and $i_{1,\mathrm{max}}$)
reduced (green curves).

At the octupole level (red curves), the asymmetries that arise in the eccentricity curve find their counterpart in the inclination
curves. Within a narrow range of angles (rightward of  $i_{1,0}|_{\mathrm{sym}}$), the orbits are allowed to flip from retrograde
to prograde. For initial inclinations bellow $i_{1,0}\sim70^\circ$, although the octupole potential introduces significant changes
in $1-e_{1,\m}$ and $i_{1,\m/\mi}$,
it is not strong enough to cause orbital flips. In accordance to what is observed in the eccentricity curves, the octupole
contributions to the left of  $i_{1,0}|_{\mathrm{sym}}$  are not affected by SRF, however, the flips observed for angles
 $i_{1,0}>i_{1,0}|_{\mathrm{sym}}$ is nearly entirely suppressed by the inclusion of these additional effects (blue curves).

%%%%%%%%%%%%%%%%%%%%%%%%
\subsubsection{Dependence on $\varepsilon_\oct$ and $\varepsilon_\extra$}

The example of Figure~\ref{fig:case1c} shows moderate differences
in maximum eccentricity and inclination range between
the quadrupole and the octupole-level solutions. To explore the behaviour
of these variables for larger octupole contributions we integrate systems
with $\varepsilon_\oct=0.042$ varying the magnitude of $\varepsilon_\mathrm{extra}$ (Table~\ref{tab:paramspace_general}).
Some of these examples are shown in Figure~\ref{fig:case2c} and \ref{fig:case2e}.

The most important difference between the top panels of Figure~\ref{fig:case2c} and Figure~\ref{fig:case1c}
is the width of the maximal eccentricity region to the right of $i_{1,0}|_{\mathrm{sym}}$
(this was already observed in the examples of Section~\ref{sec:paramspace}) and
the deepening of the high eccentricity region to the left of $i_{1,0}|_{\mathrm{sym}}$.
As $\varepsilon_\oct$ is increased, the asymmetries between the prograde
and retrograde regions of the figure become more pronounced.
Prograde orbits at intermediate inclination see their maximum eccentricities increase,
to a point that they become comparable to those seen for the retrograde orbits. It is at this
point (when $1-e_{1,\m}\lesssim10^{-4}$) that prograde orbits are allowed to flip orientations.

When SRFs are considered, the maximum eccentricities (green and blue curves) are altered
in a similar fashion as the example of Figure \ref{fig:case1c}.
The value of $e_\li$ of Equation~(\ref{eq:FTT})
is still in agreement with the global maximum of $e_{1,\m}$ (See Table 4).
On the other hand, the growth of $i_{1,\m}$ is suppressed (no flip) in prograde configurations,
while $i_{1,\mi}$ appears to vary erratically when $i_{1,0}>i_{1,0}|_{\mathrm{sym}}$
(See Figure \ref{fig:case2c}).

Figures~\ref{fig:case1e} and \ref{fig:case2e} show examples of increased $\varepsilon_\mathrm{extra}$ for
the same $\varepsilon_\oct$ as in Figures~\ref{fig:case1c} and \ref{fig:case2c}.
In these cases, $e_\mathrm{lim}$ is smaller than in  Case 2c
(Figure~\ref{fig:case2c}), which implies that SRFs are truncating $e_{1,\m}$ not only in the vicinity
of  $i_{\mathrm{tot},0}|_{\mathrm{sym}}$, but also in the lower inclination region [$i_{1,0}\in(55^\circ,75^\circ)$], thus affecting the
octupole-level effects significantly.  Despite the significant restrictions on $e_{1,\m}$ imposed by SRFs, the octupole
effects cannot be neglected for their values of $\varepsilon_\oct$, since they allow for these systems to reach eccentricities with $1-e_\li\sim5\times10^{-2}$ for
initial mutual inclinations as low as $i_{\mathrm{tot},0}\sim55^\circ$ (Figure~\ref{fig:case2e}) while the quadrupole-level calculation
would require inclinations beyond $75^\circ$ to reach similar values.

%%%%%%%%%%%%%%%%%%%%%%%%%%%%%%%%%%%%%%%%%%%%%%%%%%%%%%%%%
%%%%%%%%%%%%%%%%%%%%%%%%%%%%%%%%%%%%%%%%%%%%%%%%%%%%%%%%%
\section{Summary and conclusions}
In this paper, we have computed the extent to which energy-conserving
short-range effects alter the orbital evolution of planets and stars
in hierarchical triple systems undergoing Lidov-Kozai oscillations.
In particular, we have systematically examined how general
relativistic precession, tides and oblateness can moderate the extreme
values in eccentricity and inclination that can be achieved owing to
the octupole terms in the interaction potential.

By carrying of a sequence of numerical experiments,
we have measured the extrema in eccentricity and inclination for a variety
of hierarchical triples, systematically varying the relative strengths of
the octupole terms and of the short-range effects in terms of their contributions
to the potential energy.  The results of our calculations can be summarized
into four main findings.

(1) The importance of the octupole effects depends on the
dimensionless parameter $\varepsilon_{\oct}$ (see Equation \ref{eq:C}), which
measures the relative strength between the octupole and quadrupole
potentials. The main contribution of the octupole terms to eccentricity and inclination excitation
is limited to a range in initial inclinations or ``window of influence",
the width of which grows with $\varepsilon_{\oct}$.
As $\varepsilon_{\oct}$ decreases, the window of
influence becomes increasingly confined to
mutual inclinations close to $90^\circ$. For example,
at $\varepsilon_{\oct}\sim 0.002$, the octupole
terms are important only within a few degrees around $i_{tot,0}=90^\circ$
(see Tables 1 and 2; also see Fig.~\ref{fig:case5}).

(2) We find that short range forces can indeed
compete with the octupole-level terms in the potential, and that these additional
effects impose a strict upper limit on the maximum achievable eccentricity.
Most importantly, we find that to a very good approximation, this maximum
eccentricity can be derived analytically using the quadrupole
approximation in the test-particle limit
(see section 3.3 and Equation \ref{eq:FTT}).
This analytic limiting eccentricity
holds even for a strong octupole contribution as well as in
the general case of three comparable masses.

(3) Our results indicate that, despite the upper limit in
eccentricity (which is independent on the octupole strength), the
width of the window of influence of the octupole potential (see point 1 above)
is largely unaffected by the SRFs.

(4) We find that orbital flips are affected by the SRFs.
With increasing strength of the SRFs (characterized by the
dimensionless parameters; see Equations \ref{eq:R11}, \ref{eq:R12} and \ref{eq:R13}), orbital flips are
increasingly confined to the region close to $i_{\rm tot,0}=90^\circ$
(see Figs.~\ref{fig:case10a}\---\ref{fig:case10c} and Tables 1-2).

%%%%%%%%%%%%%%%%%%%%%%%%%%%%%%%%%%%%%%%%%%%%%%%
%%%%%%%%%%%%%%%%%%%%%%%%%%%%%%%%%%%%%%%%%%%%%%%
%%%%%%%%%%%%%%%%%%%%%%%%%%%%%%%%%%%%%%%%%%%%%%%
\section*{Acknowledgments}
This work has been supported in part by NSF grant AST-1211061, and NASA grants
NNX12AF85G, NNX14AG94G and NNX14AP31G.
DJM thanks Boaz Katz and Crist\'obal Petrovich for helpful discussions.
BL gratefully acknowledges support from the China Scholarship Council.

%%%%%%%%%%%%%%%%%%%%%%%%%%%%%%%%%%%%%%%%%%%%%%%
%%%%%%%%%%%%%%%%%%%%%%%%%%%%%%%%%%%%%%%%%%%%%%%
%%%%%%%%%%%%%%%%%%%%%%%%%%%%%%%%%%%%%%%%%%%%%%%

%%%%%%%%%%%%%%%%%%%%%%%%%%%%%%%%%%%%%%%%%%%%%%%
%%%%%%%%%%%%%%%%%%%%%%%%%%%%%%%%%%%%%%%%%%%%%%%
%%%%%%%%%%%%%%%%%%%%%%%%%%%%%%%%%%%%%%%%%%%%%%%
\appendix

\section[]{Full dynamic equations}\label{sec:Full}

We present the complete secular equations for octupole order in this section
obtained from applying the matrix projection (Equations~\ref{eq:proj} and~\ref{eq:proe})
to the vector-form equations of motion (Equations~\ref{eq:j1vec}--\ref{eq:e2vec}). 
Omitting the laborious matrix manipulation, we get:
\begin{equation}\label{eq:fulle1}
\begin{split}
\frac{de_1}{dt}&=\frac{\sqrt{1-\e1}}{64~t_K}\Bigg\{120e_1\sin^2\I\sin2\omega_1\\
&+\frac{15\varepsilon_{\oct}}{8}\cos\omega_2\Big[(4+3\e1)(3+5\cos2\I)\sin\omega_1\\
&+210\e1\sin^2\I\sin3\omega_1\Big]\\
&-\frac{15\varepsilon_{\oct}}{4}\cos\I\cos\omega_1\Big[15(2+5\e1)\cos2\I\\
&+7(30\e1\cos2\omega_1\sin^2\I-2-9\e1)\Big]\sin\omega_2\frac{}{}\Bigg\}~,
\end{split}
\end{equation}
and
\begin{equation}
\begin{split}
\frac{de_2}{dt}&=\frac{15e_1L_1\sqrt{1-e_2^2}~\varepsilon_{\oct}}{256~t_K~e_2~L_2}
\Bigg\{\cos\omega_1\Big[6-13\e1\\
&+5(2+5\e1)\cos2\I+70\e1\cos2\omega_1\sin^2\I\Big]\\
&\times\sin\omega_2-\cos\I\cos\omega_2\Big[5(6+\e1)\cos2\I\\
&+7(10\e1\cos2\omega_1\sin^2\I-2+\e1)\Big]\sin\omega_1\Bigg\}~.
\end{split}
\end{equation}
for the inner and outer eccentricities.

The time evolution of the inclinations are described by
\begin{equation}
\begin{split}
\frac{di_1}{dt}&=\frac{-~3e_1}{32~t_K~\sqrt{1-\e1}}\Bigg\{
10\sin2\I\bigg[e_1\sin2\omega_1\\
&+\frac{5\varepsilon_{\oct}}{8}(2+5\e1+7\e1\cos2\omega_1)
\cos\omega_2\sin\omega_1\bigg]\\
&+\frac{5\varepsilon_{\oct}}{8}\cos\omega_1\Big[26+37\e1-35\e1\cos2\omega_1\\
&-15\cos2\I(7\e1\cos2\omega_1-2-5\e1)\Big]\sin\I\sin\omega_2\Bigg\}~,
\end{split}
\end{equation}
and
\begin{equation}
\begin{split}
\frac{di_2}{dt}&=\frac{-~3e_1L_1}{32~t_K~\sqrt{1-e_2^2}~L_2}\Bigg\{
10\bigg[2e_1\sin\I\sin2\omega_1\\
&+\frac{5\varepsilon_{\oct}}{8}\cos\omega_1(2+5\e1-7\e1\cos2\omega_1)\sin2\I\sin\omega_2\bigg]\\
&+\frac{5\varepsilon_{\oct}}{8}\Big[26+107\e1+5(6+\e1)\cos2\I\\
&-35\e1(\cos2\I-5)\cos2\omega_1\Big]\cos\omega_2\sin\I\sin\omega_1\Bigg\}~.
\end{split}
\end{equation}
We also write the longitudes of ascending nodes as a function of time
\begin{equation}
\begin{split}
\frac{d\Omega_1}{dt}&=\frac{d\Omega_2}{dt}=\frac{-3\csc i_1}{32~t_K~\sqrt{1-\e1}}
\Bigg\{2\bigg[(2+3\e1-5\e1\cos2\omega_1)\\
&+\frac{25\varepsilon_{\oct}e_1}{8}\cos\omega_1(2+5\e1-7\e1\cos2\omega_1)\cos\omega_2\bigg]\\
&\times\sin2\I-\frac{5\varepsilon_{\oct}e_1}{8}\Big[35\e1(1+3\cos2\I)\cos2\omega_1\\
&-46-17\e1-15(6+\e1)\cos2\I\Big]\sin\I\sin\omega_1\sin\omega_2\Bigg\}~.
\end{split}
\end{equation}
Finally, the argument of periapse for the inner and outer binaries evolve according to
\begin{equation}\label{eq:domega_dt}
\begin{split}
\frac{d\omega_1}{dt}&=\frac{3}{8~t_K}\Bigg\{\frac{1}{\sqrt{1-\e1}}\Big[4\cos^2\I+(5\cos2\omega_1-1)\\
&\times(1-\e1-\cos^2\I)\Big]+\frac{L_1\cos\I}{L_2\sqrt{1-e_2^2}}\Big[2+\e1(3\\
&-5\cos2\omega_1)\Big]\Bigg\}+\frac{15\varepsilon_{\oct}}{64~t_K}
\Bigg\{\left(\frac{L_1}{L_2\sqrt{1-e_2^2}}+\frac{\cos\I}{\sqrt{1-\e1}}\right)\\
&\times e_1\bigg[\sin\omega_1\sin\omega_2\Big[10(3\cos^2\I-1)(1-\e1)+A\Big]\\
&-5B\cos\I\cos\Theta\bigg]-\frac{\sqrt{1-\e1}}{e_1}\Big[10\sin\omega_1\sin\omega_2\cos\I\\
&\times\sin^2\I(1-3\e1)+\cos\Theta(3A-10\cos^2\I+2)\Big]\Bigg\}~,
\end{split}
\end{equation}
and
\begin{equation}\label{eq:fullomega2}
\begin{split}
\frac{d\omega_2}{dt}&=\frac{3}{16~t_K}\Bigg\{\frac{2\cos\I}{\sqrt{1-\e1}}
\Big[2+\e1(3-5\cos2\omega_1)\Big]\\
&+\frac{L_1}{L_2\sqrt{1-e_2^2}}\bigg[\frac{}{}4+6\e1+(5\cos^2\I-3)\\
&\times\Big[\frac{}{}2+\e1(3-5\cos2\omega_1)\Big]\bigg]\Bigg\}
-\frac{15\varepsilon_{\oct}e_1}{64~t_K~e_2}\\
&\times\Bigg\{\sin\omega_1\sin\omega_2\bigg[
\frac{L_1(4e_2^2+1)}{e_2~L_2\sqrt{1-e_2^2}}10\cos\I\sin^2\I\\
&\times(1-\e1)-e_2\left(\frac{1}{\sqrt{1-\e1}}+\frac{L_1\cos\I}{L_2\sqrt{1-e_2^2}}\right)\\
&\times\Big[A+10(3\cos^2\I-1)(1-\e1)\Big]\bigg]+\cos\Theta\\
&\times\bigg[5B\cos\I e_2\left(\frac{1}{\sqrt{1-\e1}}+\frac{L_1\cos\I}{L_2\sqrt{1-e_2^2}}\right)\\
&+\frac{L_1(4e_2^2+1)}{e_2~L_2\sqrt{1-e_2^2}}A\bigg]\Bigg\}~,
\end{split}
\end{equation}
where we define
\begin{equation}
A\equiv4+3\e1-\frac{5}{2}B\sin^2\I~,~~~\\
B\equiv2+5\e1-7\e1\cos2\omega_1~,
\end{equation}
and
\begin{equation}\label{eq:fullcos}
\cos\Theta\equiv-\cos\omega_1\cos\omega_2-\cos\I\sin\omega_1\sin\omega_2~.
\end{equation}

\label{lastpage}

\end{document}